\documentclass[final,3p,times]{elsarticle}

\usepackage{amsfonts}
\usepackage{color}
\usepackage{bm}
\usepackage{graphicx}
\usepackage{epstopdf}
\usepackage{graphicx, subfigure}

\usepackage{amssymb}
\usepackage{amsthm}
\usepackage{amsmath}

\biboptions{comma,square,sort&compress}

\journal{Comput. Methods Appl. Mech. Engrg.}

\begin{document}

\begin{frontmatter}



\title{A Spectral Multiscale Method for Wave Propagation Analysis: Atomistic-Continuum Coupled Simulation}



\author[label1]{Amit K. Patra}
\ead{ssamitkpatra@aero.iisc.ernet.in}
\author[label1]{S. Gopalakrishnan}
\cortext[cor1]{Corresponding author.}
\ead{krishnan@aero.iisc.ernet.in}
\author{Ranjan Ganguli\corref{cor1}\fnref{label1}}
\ead{ganguli@aero.iisc.ernet.in}

\address[label1]{Department of Aerospace Engineering, Indian Institute of Science, Bangalore 560012, India}

\begin{abstract}
In this paper, we present a new multiscale method which is capable of coupling atomistic and continuum domains for high frequency wave propagation analysis. The problem of non-physical wave reflection, which occurs due to the change in system description across the interface between two scales, can be satisfactorily overcome by the proposed method. We propose an efficient spectral domain decomposition of the total fine scale displacement along with a potent macroscale equation in the Laplace domain to eliminate the spurious interfacial reflection. We use Laplace transform based spectral finite element method to model the macroscale, which provides the optimum approximations for required dynamic responses of the outer atoms of the simulated microscale region very accurately. This new method shows excellent agreement between the proposed multiscale model and the full molecular dynamics (MD) results. Numerical experiments of wave propagation in a 1D harmonic lattice, a 1D lattice with Lennard-Jones potential, a 2D square Bravais lattice, and a 2D triangular lattice with microcrack demonstrate the accuracy and the robustness of the method. In addition, under certain conditions, this method can simulate complex dynamics of crystalline solids involving different spatial and/or temporal scales with sufficient accuracy and efficiency.
\end{abstract}


\end{frontmatter}


\section{Introduction}
The mechanics of matter in almost every part of science is governed by physical laws of different character at different spatial and/or temporal scales. Unfortunately, there is no single theory at present which is applicable for all the scales. The rapidly developing applied science and engineering fields demand detailed information of physical phenomena at different length and
time scales, even at atomistic scales. Continuum mechanics alone cannot describe these phenomena, since the continuum approximation at these small scales becomes questionable. Again, classical atomistic simulation alone is not applicable directly for a realistic physical system due to limitations in the computational capabilities available. Due to the computational limitation at present, a very small portion of the whole domain can be modeled as fine scale (microscale) considering physics at the smallest scale of interest. However, multiscale analysis aims to overcome this snag by coupling the atomistic and continuum description in a single computational framework. The multiscale approach aims at incorporating models of the sub-domain governed by the physics at small scale and then coupling these to the continuum model of the whole domain using matching interfacial condition. These matching conditions must resist spurious reflection of the elastic wave along the interface of the two computational domains with incompatible dispersive characteristics, and inadequate meshes.

Over the last few years, enormous efforts were made in designing multiscale algorithms and their successful application in various physical systems is widely reported. Kohlhoff et al. \cite{Kohlhoff} used combined finite element and atomistic method to study the crack propagation in BCC crystal. Abraham et al. \cite{Abraham} developed the domain decomposition based (DD) ``macroscopic, atomistic, ab initio dynamics'' method (MAAD) to simulate brittle fracture. Several bridging domain methods were developed thereafter \cite{Broughton, RuddCGM, Rudd, RuddCGM2, Belytschko} for multiscale materials.  A handshaking region is used to minimize interfacial reflection in most of these earlier multiscale methods \cite{Abraham, Broughton, Rudd}. Basu et al. \cite{Basu} used a damping term to absorb numerical reflection in their perfectly matched layer method. Unfortunately, all these methods are limited for general applicability as they did not allow accurate passage of fine scale fluctuation across the interface.

The powerful quasicontinuum  method (QC), originally developed by Tadmor et al. \cite{Tadmor}, uses the Cauchy-Born rule to link atomistic level to continuum level and representative atoms with adaptively refined mesh. Finite element mesh is gradually refined to atomic scale near the crack or dislocation to reduce interfacial reflection and the local energy of the atomistic region is obtained by an energy summation rule. This method has been successfully applied to simulate dislocation motion \cite{Tadmor, Tadmor2, Shenoy}, nano-indentation \cite{Tadmor2}, and fracture \cite{Miller, Shenoy, MLuskin} of solids at zero temperature mainly in static and quasi-static condition. Dupuy et al. \cite{LMDTadmor} developed the finite-temperature quasicontinuum method. Many other researchers \cite{ZTangAluru, JMnGVenturini} extended the QC method separately for finite-temperature simulation. However, this method is not appropriate for high frequency wave propagation analysis.

E and Engquist \cite{WE2002} developed heterogeneous multiscale method (HMM) for coupling atomistic-continuum dynamic simulation. Using this HMM framework, Li et al. \cite{Xli2005} proposed a new multiscale method for modeling dynamics of solids at finite temperature. The main philosophy of their method in coupling the microscale (MD) with the macroscale is to derive the microscale equation in form of conservation laws of mass, momentum and energy, since the macroscale model is a system obeying conservation laws. A series of papers, published thereafter illustrated the same methodology \cite{WE2007, Xli2007, Xli2010}. However, these methods also could not completely eliminate the spurious interfacial reflections in high frequency dynamic simulation.

Recently, the bridging scale method (BSM) was introduced by Wagner and Liu \cite{WagnerBSM}. Soon after, a series of papers \cite{ParkMS, Liu, TangBSM, TangPSBSM, Farrell, Qian} showcased the same framework in modeling multiscale dynamic problems. They consider a domain decomposition of the time-domain fine scale displacement field into a mean part and a fine fluctuation part. For decomposition of total fine scale displacement field, they use a linear projection operator, which cannot effectively minimize the energy transition due to the fine fluctuation across the interface. Thus, this approach is not able to resist the interfacial non-physical wave reflection completely.

It is understood that an efficient domain decomposition of the fine scale displacement field along with a potent coarse scale continuum model can substantially reduce this interfacial spurious wave reflection. Modeling the coarse scale in the frequency domain is a superior alternative. Chatfield \cite{Chatfield} stated that spectral analysis in frequency domain is the synthesis of waveforms through the superposition of many frequency components. Doyle \cite{Doyle, DoyleTN, DoyleBook} recast this concept into a matrix based methodology known as spectral finite element method (SFEM). Subsequently, many other researchers \cite{Gopalakrishnan, Gopalakrishnan2, Deepak, Vinod, GKrishnanBook} have successfully applied and contributed to the SFEM. As in the conventional FEM, in SFEM, the dynamic stiffness matrix can be derived directly from the weak form of the governing equations in the frequency domain \cite{Gopalakrishnan}. In SFEM, there is no generation of separate mass matrix as in conventional FEM, however it is built-in into the exact dynamic stiffness matrix. In this method, the input excitation in time domain is transformed into the frequency domain image by forward fast fourier transform (FFT) and the dynamic equation is solved for each frequency for the nodal variable of interest. Then the nodal solutions obtained for all frequencies are transformed back to time domain by using inverse FFT. The main drawback of this Fourier transform based SFEM approach is that it cannot effectively handle waveguides of short lengths. To address this lacuna, Mitra and Gopalakrishnan \cite{MMitra} developed wavelet transform based spectral FEM, which can handle short waveguides with ease due to the use of compactly supported Daubechies wavelets in their formulation. Although wavelets provide good time resolution, their frequency resolution is quite poor. Beyond certain fraction of Nyquist frequency, they introduce a non-existent spurious dispersion \cite{SGBook}. Recently, Igawa et al. \cite{Igawa} developed Laplace transform based spectral FEM for wave propagation analysis of finite length waveguide. Some other researchers \cite{Blais, Kishor, Murthy} have applied this new method in wave propagation analysis of several finite domain mediums.

In this work, we present a new frequency domain based spectral multiscale method (SMM) which is capable of modeling atomistic-continuum simulation for high frequency wave propagation problems with excellent accuracy. The promising feature of SMM is the development of a powerful coarse scale model. We propose the domain decomposition of the fine scale spectral displacement field and model the coarse scale with Laplace transform based spectral FEM (NLSFEM). Therefore, we decompose the spectral (Laplace domain) fine scale displacement field into a mean part and a fine fluctuation part. The mean part of the spectral displacements at fine scale grid can be easily obtained from the coarse grid displacements via dynamic shape functions, which, instead of being simple polynomials, are exact displacement distributions. Fine fluctuations, which are caused mostly by higher order modes, are almost eliminated by this approach. This in turn significantly reduces the spurious numerical wave reflection at the interface. In a single computation, we use NLSFEM for the coarse scale and time domain MD/FEM for the fine scale analysis. Derivation of the efficient and accurate coarse grid equations, which is vital for a multiscale algorithm, is easily viable in the spectral domain for many realistic linear elastic systems. In modeling the interface, we do not use any handshaking region. It is noticed that internal force on interfacial atoms/nodes in the fine scale sub-domain involve the displacements of some atoms/nodes just outside the simulated fine scale region, called ghost point atoms/nodes. Displacements at these atoms/nodes are reconstructed by summing up the mean displacements obtained from the coarse grid displacements and the fine fluctuations part which is computed directly from the equation derived for fine fluctuations. In the present approach, the powerful coarse scale model maximizes the energy associated with the mean displacement, which substantially reduces the error in computations of fine fluctuations through time history convolution. This is a major source of numerical error in most of the earlier domain decomposition (DD) methods \cite{WagnerBSM, ParkMS, Liu, TangBSM, TangPSBSM, Farrell, Qian}. In addition, the capability of a huge reduction in system size makes our SMM generally applicable and computationally economic.

\subsection{Outline}
The rest of this paper is organized as follows. First, the formulation of the spectral multiscale method including derivation of the coarse grid equation and interfacial conditions at zero temperature is presented in section $2$. In section $3$, the continuum approximation of atomistic equations and NLSFEM Formulations for 1D and 2D atomistic systems is described. The numerical Laplace transformation technique is described in section $4$.  In section $5$, several numerical schemes involved and an algorithm for the proposed method is presented. In section $6$, the SMM is applied to wave propagation problems in a 1D harmonic lattice, a 1D nonlinear lattice with next nearest neighbor interaction, a 2D square Bravais lattice, and a 2D hexagonal close-packed lattice with microcrack. In section $7$, a discussion of the numerical results is presented. Concluding remarks are made in the final section $8$.

\section{General Formulation of Atomistic-Continuum Model}

In the spectral multiscale computations, the atomistic-continuum coupled models are considered. Here, the atomistic dynamics of a material system in $\Omega \subset \mathbb{R}^{3}$, consisting of $n_a$ atoms is considered. The initial position of the $i^{th}$ atom of the system when at rest is $x_{i0}$ and the displacement of the atom is given by $u_i = x_i - x_{i0}$. Here, the displacements $u_i$ and the actual position $x_i$ are functions of time. Motion of the atomistic system obeys the Newton's second law \cite{WagnerBSM, TangBSM, Frenkel} as
\begin{equation}\label{eq1}
 [M_{f}]\{\ddot{u}\}=\{f_{int}\}+\{f_{ext}\}
\end{equation}
 where $\{u\}, \{f_{int}\}, \{f_{ext}\} \in \mathbb{R}^{3n_a}$ are the displacement, internal force and external force, respectively. The mass matrix is $ [M_f]=diag(m_1I_{3\times3},...,m_{n_a}I_{3\times3})$ and $m_i>0$ for all $i$. The internal force here is due to the interatomic interactions. The interatomic potential of the material system is a function of atomic displacements with respect to its frame of reference. The internal force is related to the interatomic potential by the relation
\begin{equation}\label{eq2}
 \{f_{int}\}= -\nabla_u U(u)
\end{equation}
The spectral (in Laplace domain) form of the equation of motion of each atom location, under same initial and boundary conditions, can be written as
\begin{equation}\label{eq3}
 [\hat{K_{f}}]\{\hat{u}\} = \{\hat{f}_{ext}\} + \{\hat{\beta}(s)\}
\end{equation}
where $\{\hat{u}\}, \{\hat{f_{ext}}\} \in \mathbb{C}^{3n_a}$ represents the displacement, internal force and external force in the Laplace domain, respectively. The complex stiffness matrix $[\hat{K_f}]$ is a function of the Laplace variable $s$ and represents the exact dynamic stiffness matrix of the atomistic system. The initial condition is taken into account by the vector $\{\hat{\beta}(s)\}$. Depending upon the nature of interatomic potential $U$, $\{f_{int}\}$ may become a nonlinear function of $\{u\}$. We do not consider the nonlinearity in the spectral form of equation of motion in the present formulation.

A cogent and consistent formulation to show the mathematical basis of the new spectral multiscale method have been derived in the following subsections. In subsection $2.1$, the decomposition of displacement field and development of the coarse grid equation in spectral domain is derived. An interfacial condition for the spectral multiscale method is presented in subsection $2.2$.

\subsection{Decomposition of displacement field and development of coarse grid equation}

\begin{figure}[!t]
\centering
\includegraphics[scale=0.5]{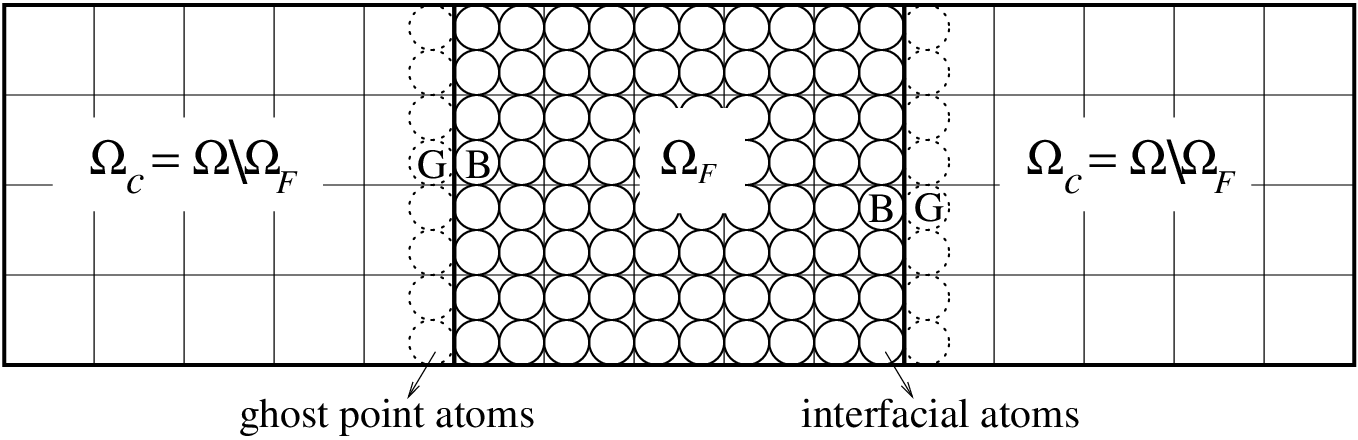}
\caption{The schematic of the whole domain $\Omega$ and the fine scale domain $\Omega_F$ considered in the spectral multiscale method. The region $\Omega_C$ represents domain only used for coarse scale analysis. The positioning of ghost (G) and boundary (B) point atoms/nodes along the interface is shown.}
\label{mdfedomain}
\end{figure}

Now we separate the whole physical domain into two sub-domains, namely $\Omega_F$ and $\Omega_C = \Omega\setminus\Omega_F$ (Fig. \ref{mdfedomain}). In the whole domain $\Omega $, we decompose the fine spectral displacement $\hat{u}$ into a mean part $\bar{\hat{u}}$, and a fine fluctuation part $\breve{\hat{u}}$ as given below.
\begin{equation}\label{eq4}
\{\hat{u}\}= \{\bar{\hat{u}}\} + \{\breve{\hat{u}}\}
\end{equation}
The focus here is to perform Molecular Dynamics(MD) computations solely in $\Omega_F$ to accurately capture the non-linear dynamics. To continue with the proper fine scale computations only in sub-domain $\Omega_F$, displacement response of atoms (ghost point atoms) just outside the fine scale sub-domain needs to be known. The mean part of these displacement response of ghost point atoms can be obtained from the coarse grid displacement via some interpolation function.
Therefore, we consider a coarse grid of total $n_c$ nodes over the whole domain $\Omega$ including the regions $\Omega_F$ where MD computations will also be performed. We assign a spectral displacement $\hat{d_j}$ at each $j^{th}, (j=1,...,n_c)$ coarse grid node. The coarse grid displacement vector is $\{\hat{d}\}$ and macroscopic computations are performed over the whole domain $\Omega$ for $\{\hat{d}\}$ in spectral domain. Assuming fine fluctuations in $\Omega_c$ are negligible, we express the motion only in coarse scale. This restriction substantially reduces the computing load and memory requirements, compared to full MD simulation.

The mean fine displacement $\{\bar{\hat{u}}\}$ is related to coarse grid displacement $\{\hat{d}\}$ by
\begin{equation}\label{eq5}
 \{\bar{\hat{u}}\} =\hat{[N]}\{{\hat{d}}\}
\end{equation}
where $\hat{[N]}$ is an interpolation matrix of size $3n_a\times3n_c$ consisting of spectral dynamic shape functions \cite{DoyleBook, GKrishnanBook} or other shape function. We already assigned $n_a$ and $n_c$ as the number of atoms and the number of coarse grid nodes, respectively. Our objective here is to minimize the value of fine fluctuation $\{\breve{\hat{u}}\}$ to obtain optimal approximation of $\{\hat{u}\}$. For this, we minimize the energy residual corresponding to $\{\breve{\hat{u}}\}$ given by
\begin{equation}\label{eq6}
 R=(\{\hat{u}\} - \hat{[N]}\{{\hat{d}}\})^T\hat{[K_f]}(\{\hat{u}\} - \hat{[N]}\{{\hat{d}}\})
\end{equation}
Solving the above equation for $\{\hat{d}\}$, we get
\begin{equation}\label{eq7}
\{\hat{d}\}=[\hat{K_c}]^{-1}[\hat{N}]^{T}[\hat{K_f}]\{\hat{u}\}
\end{equation}
Here, $[\hat{K_c}]=[\hat{N}]^{T}[\hat{K_f}][\hat{N}]$ is the effective spectral stiffness matrix of rank ${3n_c}$. $[\hat{K_c}]$ is symmetric and is viewed as the spectral stiffness matrix of coarse grid equation. Using Eqs. (\ref{eq4}), (\ref{eq5}) and (\ref{eq7}); the coarse-fine decomposition is defined as
\begin{equation}\label{eq8}
\{\bar{\hat{u}}\}=\hat{[P]}\{{\hat{u}}\}, \; \;   \{\breve{\hat{u}}\}=\hat{[Q]}\{{\hat{u}}\}
\end{equation}
where, $\hat{[P]}=[\hat{N}][\hat{K_c}]^{-1}[\hat{N}]^{T}[\hat{K_f}]$ is the projection operator and $\hat{[Q]}=[I]-\hat{[P]}$ is the complementary projection operator.

The equation of coarse grid is obtained by pre-multiplying equation (\ref{eq3}) by $[\hat{K_c}]^{-1}[\hat{N}]^{T}$
\begin{equation}\label{eq9}
\{\hat{d}\}=[\hat{K_c}]^{-1}[\hat{N}]^{T}(\{\hat{f}_{ext}\} + \{\hat{\beta}(s)\})
\end{equation}
Or, the equation (\ref{eq9}) can be satisfactorily replaced by the continuum equation in Laplace domain as
\begin{equation}\label{eq10}
\{\hat{d}\}=[\hat{K_s}]^{-1}(\{\hat{F}_{ext}\} + \{\hat{\Gamma}(s)\})
\end{equation}
In our coarse scale computations, instead of $[\hat{K_c}]$, we use exact spectral stiffness matrix $[\hat{K_S}]$ to avoid the extra load in the computations of projection operators and thus, for $[\hat{K_c}]$. Therefore, we derive the continuum equation of motion incorporating the physics at fine scale (section $3$) which gives $[\hat{K_S}]$ as the best approximation of $[\hat{K_c}]$.

\subsection{Development of interfacial condition}
Since details of the wave propagating in atomistic model cannot be represented exactly at the coarse scale (continuum) level, special treatment is required at the coarse-fine interface to avoid non-physical numerical reflection. Again, the matching between the coarse-fine interface should be such that it allows accurate information transmission on either sides of the interface. In
order to minimize this non-physical reflection, several coupling approaches have been explored recently \cite{Adelman, Cai, EH, EH2, KarpovBC, WagnerMDBC, STangBC, Xli2006, Xli2007}. Adelman and Doll \cite{Adelman} are the first researchers to model the matching interfacial condition for a coupled domain. They included a time history integral over atomic velocities convolved with a damping kernel matrix function $\beta(t)$ in their reduced equation of motion. However, in most practical cases, it is difficult to find $\beta(t)$ in closed form. Cai et al. \cite{Cai} numerically solved for $\beta(t)$ for a coupled molecular dynamics (MD) simulation. E and Huang \cite{EH, EH2} avoided the time history integral and used a truncated discrete summation whose coefficients are obtained by optimizing for minimum internal reflection at the atomistic/continuum interface in multiscale modeling of crystal. However, both of these methods are limited for general applicability. Wagner et al. \cite{WagnerMDBC} proposed a method for finding $\beta(t)$, which is based on a linearization in the vicinity of the boundary. Karpov et al. \cite{KarpovBC} used a memory function, related to the lattice dynamics Green's function, to represent the harmonic response of outer, non-simulated region. Tang et al. \cite{STangBC} used a finite difference approach with velocity interfacial condition for multiscale computations of crystalline solids with relatively strong nonlinearity and large deformation. Li and E \cite{Xli2006, Xli2007} proposed a variational formalism to construct boundary conditions that minimize total phonon reflection. Another interesting work is done in this field by Gonella and Ruzzene \cite{Gonella}. To the best of the author's knowledge, all the approaches reported in the literatures cannot overcome the snag of the spurious numerical reflection completely.

In the present work, we focus only on obtaining fine fluctuations of the ghost point atoms considering harmonic lattice. The interatomic force is assumed to be linear function of displacements and can be written as $ f_{int}(u)=-Ku$. Here, $K$ is the stiffness matrix of the lattice. The equation of motion (\ref{eq1}) can be written as
\begin{equation}\label{GLE1}
 [M_f]\{\ddot{u}\}=-[K]\{u\}+\{f_{ext}\}
\end{equation}
Using Laplace transformation Eq. (\ref{GLE1}) is transformed into the spectral domain; we get the conventional dynamic equation as
\begin{equation}\label{GLE2}
 (s^2[M_f]+[K])\{\hat{u}(s)\}=\{\hat{f_{ext}}\} + s[M_f]\{{u}\}|_{t=0} + [M_f]\{\dot{u}\}|_{t=0}
\end{equation}
Or,
\begin{equation}\label{GLE3}
 [\hat{K}_{eff}]\{\hat{u}\}=\{\hat{f_{ext}}\} + \{\hat{\beta}(s)\}
\end{equation}
where, $s$ is the Laplace state variable and the complex matrix $[\hat{K}_{eff}]$ is the dynamic stiffness matrix for the fine scale. The extra forcing vector $\hat{\beta}(s)=s[M_f]\{{u}\}|_{t=0} + [M_f]\{\dot{u}\}|_{t=0}$ takes the initial fine scale energy into account. Pre-multiplying Eq. (\ref{GLE1}) by $[M_f]^{-1}$ and defining $A=[M_f]^{-1}[K]$ and $q=[M_f]^{-1}\{f_{ext}\}$, we decompose the equation into two parts from $\Omega_F$ and $\Omega_C$ as
\begin{equation}\label{GLE5}
\left(
  \begin{array}{c}
    \ddot{u}_F \\
    \ddot{u}_C \\
  \end{array}
\right)=-\left[
          \begin{array}{cc}
            A_{FF} & A_{FC}\\
            A_{CF} & A_{CC}\\
          \end{array}
         \right]\left(
                 \begin{array}{c}
                   u_F\\
                   u_C\\
                 \end{array}
                \right) + \left(
                           \begin{array}{c}
                             {q}_F \\
                             {q}_C \\
                           \end{array}
                          \right)
\end{equation}
Then the equations for $\ddot{u}_F$ in (\ref{GLE5}) can be written as
\begin{align}\label{GLE6}
 \{\ddot{u}_F\}  = - A_{FF}\{{u}_F\} - A_{FC}\{{u}_C \}+ \{{q}_F\}
\end{align}

Therefore, for the fine scale computations of $\ddot{u}_F$ in $\Omega_F$ domain, displacement time history of few ghost point atoms/nodes in $\Omega_C$ (i.e.,$\{{u}_C\}$) is required to be known. These displacements can be obtained by transforming spectral displacements $\{\hat{u}_C\}$ computed in $\Omega_C$ to time domain. The mean part $\{\bar{\hat{u}}_C\}$ of the total spectral displacements $\{\hat{u}_C\}= \{\bar{\hat{u}}_C\} + \{\breve{\hat{u}}_C\}$ in $\Omega_C$ is obtained from coarse grid displacements (\ref{eq10}) using dynamic shape functions (\ref{eq5}). For fine fluctuation part $\{\breve{\hat{u}}_C\}$, we proceed as follows. Using complementary projection operator we can show
\begin{equation}\label{GLE7}
[\hat{Q}]^{T}\hat{[K_f]}=\hat{[K_f]}-\hat{[K_f]}\hat{[N]}\hat{[K_c]}^{-1}\hat{[N]}^{T}[\hat{K}_f]=[\hat{K}_f][\hat{Q}]
\end{equation}
Therefore, pre-multiplying Eq. (\ref{GLE3}) by $\hat{[Q]}^{T}$ and using the definition $\{\breve{\hat{u}}\}=[\hat{Q}]\{{\hat{u}}\}$, we obtain the governing equation for $\{\breve{\hat{u}}\}$
\begin{equation}\label{GLE8}
[\hat{K}_f]\{\breve{\hat{u}}\} = [\hat{Q}]^{T}\left(\{\hat{f}_{ext}\} + \{\hat{\beta}\}\right) = [\hat{Q}]^{T}\{\hat{f}_{\alpha}\}
\end{equation}
Or
\begin{equation}\label{GLE9}
[\hat{K}_{CC}]\{\breve{\hat{u}}_C\}+[\hat{K}_{CF}]\{\breve{\hat{u}}_F\} = [\hat{Q}_{FC}]^{T}\{\hat{f}_{\alpha_F}\} + [\hat{Q}_{CC}]^{T}\{\hat{f}_{\alpha_C}\}
\end{equation}
And,
\begin{equation}\label{GLE10}
[\hat{K}_{FC}]\{\breve{\hat{u}}_C\}+[\hat{K}_{FF}]\{\breve{\hat{u}}_F\} = [\hat{Q}_{FF}]^{T}\{\hat{f}_{\alpha_F}\} + [\hat{Q}_{CF}]^{T}\{\hat{f}_{\alpha_C}\}
\end{equation}
From (\ref{GLE10}), we obtain
\begin{equation}\label{GLE11}
\{\breve{\hat{u}}_F\} = \hat{[K_{FF}]}^{-1}([\hat{Q}_{FF}]^{T}\{\hat{f}_{\alpha_F}\} + [\hat{Q}_{CF}]^{T}\{\hat{f}_{\alpha_C}\} - [\hat{K}_{FC}]\{\breve{\hat{u}}_C\})
\end{equation}
From (\ref{GLE9}) and (\ref{GLE11}) we obtain the equation for $\{\breve{\hat{u}}_C\}$ as
\begin{align}\label{GLE12}
\{\breve{\hat{u}}_C\}=&([\hat{K}_{CC}] - [\hat{K}_{CF}][\hat{K}_{FF}]^{-1}[\hat{K}_{FC}])^{-1}([\hat{Q}_{CC}]^{T}\{\hat{f}_{\alpha_C}\}
+ [\hat{Q}_{FC}]^{T}\{\hat{f}_{\alpha_F}\}\nonumber \\ & - [\hat{K}_{CF}][\hat{K}_{FF}]^{-1}([\hat{Q}_{CF}]^{T}\{\hat{f}_{\alpha_C}\}
+ [\hat{Q}_{FF}]^{T}\{\hat{f}_{\alpha_F}\}))
\end{align}

The equation (\ref{GLE12}) gives the exact value of fine fluctuations $\{\breve{\hat{u}}_C\}$ of atoms/nodes in $\Omega_C$ including the ghost point atoms/nodes. However, it is cumbersome to compute projection operator $\hat{[P]}$ and complementary projection operator $\hat{[Q]}$ for very high fine scale degrees of freedom. Thus, it is not economical to compute $\{\breve{\hat{u}}_C\}$ by using (\ref{GLE12}). We assume exact dynamic shape functions \cite{DoyleBook, GKrishnanBook} as interpolation functions in $\hat{[N]}$ which makes $\hat{[P]}$ almost an identity matrix and $\hat{[Q]}$ becomes a null matrix. Therefore, we can neglect the forcing part on the right-hand side of (\ref{GLE9}) and rewrite the equation as
\begin{equation}\label{GLE13}
\{\breve{\hat{u}}_C\} = [\hat{K}_{CC}]^{-1}[\hat{K}_{CF}]\{\breve{\hat{u}}_F\} = \Theta (s)\{\breve{\hat{u}}_F\}
\end{equation}
where $\Theta (s) = [\hat{K}_{CC}]^{-1}[\hat{K}_{CF}]$. Taking the inverse Laplace transform of (\ref{GLE13}), we obtain the equation of fine fluctuation in time domain as
\begin{equation}\label{GLE14}
\{\breve{u}_C(t)\} = \int_0^{t}\theta (t-\tau)\{\breve{u}_F\}(\tau)d\tau
\end{equation}
where $\theta (t)$ is the memory kernel matrix. Therefore, we have the following form of generalized Langevin equation (GLE).
\begin{equation}\label{GLE15}
\{\ddot{u}_F\}  = - A_{FF}\{{u}_F\} - A_{FC}\{\bar{u}_C\} - A_{FC}\int_0^{t}\theta (t-\tau)\{\breve{u}_F\}(\tau)d\tau + \{{q}_F\}
\end{equation}
where $\{\bar{u}_C\}= L^{-1}\{\{\bar{\hat{u}}_C\}\}$. This equation (\ref{GLE15}) resembles the form of GLE described in the literatures \cite{KarpovBC, WagnerMDBC, ParkMS, TangBSM, Farrell, Xli2010}. The Eq. (\ref{GLE14}) gives very good approximation of fine fluctuation $\{\breve{u}_C\}$ at ghost point atoms/nodes for the assumption of nearest neighbor interaction. However,  Eq. (\ref{GLE14}) can be satisfactorily applied for next nearest neighbor interaction.

\section{Continuum Approximation of Atomistic Equations and Spectral Element Formulation}
Modeling of an efficient coarse scale equation to incorporate the physics at fine scale is very important in a multiscale algorithm. This is because the solutions of the coarse scale provide the mean part of the necessary interfacial conditions for the windowed fine scale analysis. In the previous section, we had developed a method of bridging of coarse scale equation of motion in Laplace domain with the fine scale. In this section we describe how the continuum approximations can be obtained from atomistic models and then we will go on to outline the numerical Laplace transform based spectral finite element (NLSFEM) formulation from these continuum equations.
\begin{figure}[!h]
\centering
\includegraphics[scale=0.5]{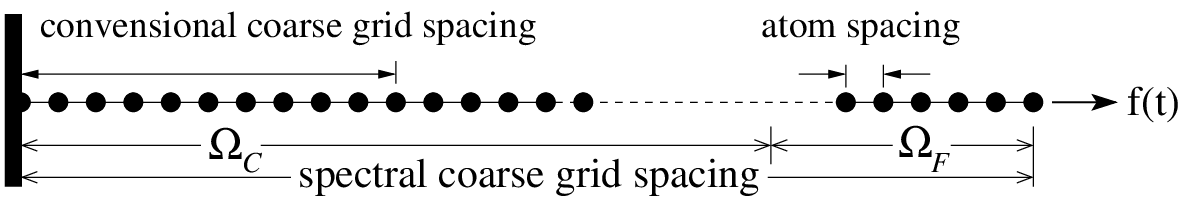}
\caption{$1$D chain of atoms used in examples $6.1$ and $6.2$.}
\label{harmoniclattice}
\end{figure}

\subsection{Spectral element formulation for axial wave propagation in 1D harmonic and nonlinear lattices}
We consider a simple 1D harmonic lattice model (Fig. \ref{harmoniclattice}) for modeling longitudinal wave propagation. The lattice consists of $n$ identical atoms of equal mass $m_a$ connected by $n$ identical, massless, lossless, perfectly linear springs of stiffness $k_a$. We assume all atoms are initially at rest and the position of the $i^{th}$ atom is given by
\begin{equation}\label{SERE01}
x_i=il_a; \;\;  \;\; i=0,1,2,3...,n
\end{equation}
Here, $l_a$ is the equilibrium atomic spacing. Assuming the displacements are small compared to equilibrium interatomic spacing, the homogeneous equation of motion of the $j^{th}$ atom can be described as \cite{Kittel}
\begin{equation}\label{SERE02}
m_a\ddot{u}_j = k_a(u_{j+1} - 2u_j + u_{j-1})
\end{equation}
Considering the traveling wave solution of the type $u_j=A e^{i(k_ljl_a-\omega t)}$, the dispersion relation and the expression group velocity $C_g$ are given by \cite{Kittel}
\begin{align}\label{SERE031}
 \omega^2= 2\frac{k_a}{m_a}(1 - \cos k_ll_a)\nonumber\\
 C_g = \frac{d\omega}{dk_l} = l_a\sqrt{\frac{k_a}{m_a}} \cos \frac{k_ll_a}{2}
 \end{align}
where $k_l$ and $\omega$ are longitudinal wave number and circular frequency, respectively. Now assuming the 1D chain of atoms as a rod of uniform cross-section $A$ with Young's modulus $E$ and density $\rho$, letting $m_a=\rho A l_a$ and $k_a = EA/l_a$, the equation (\ref{SERE02}) can be rewritten as
\begin{equation}\label{SERE03}
\rho \frac{\partial^2 u_j}{\partial t^2} = E\frac{(u_{j+1} - 2u_j + u_{j-1})}{{l_a}^2}
\end{equation}
The equation of motion (\ref{SERE03}) can be reduced to the continuum limit equation for the displacement field $u$, considering $l_a\rightarrow 0$ as \cite{DoyleBook, GKrishnanBook, Kittel}
\begin{equation}\label{SERE1}
E\frac{\partial^2 u(x,t)}{\partial x^2} - \rho \frac{\partial^2 u(x,t)}{\partial t^2}=0
\end{equation}
where, $u(x,t)$ is the axial displacement of the chain. Considering $l_a\rightarrow 0$, the limiting value of the group velocity $C_g$ in the atomic chain is given by \cite{Kittel}
\begin{equation}\label{SERE2}
C_g = \sqrt{\frac{E}{\rho}}
\end{equation}
Transforming Eq. (\ref{SERE1}) to Laplace domain yields
\begin{equation}\label{SERE3}
\frac{\partial^2 \hat{u}(x,s)}{\partial x^2} - s^2 \frac{\rho}{E} \hat{u}(x,s) = - s \frac{\rho}{E} {u}(x,t)|_{t=0} -  \frac{\rho}{E} \frac{\partial {u}(x,t)}{\partial t}|_{t=0}
\end{equation}
The homogeneous general solution of Eq. (\ref{SERE3}) considering any length $L$ of the chain can be written as
\begin{equation}\label{SERE4}
\hat{u}(x,s) = \texttt{c$_1$}e^{-ik_lx} + \texttt{c$_2$}e^{-ik_l(L-x)}
\end{equation}
where, $\texttt{c$_1$}$ and$\texttt{c$_2$}$ are arbitrary constants and $k_l$ is expressed as
\begin{equation}\label{SERE5}
k_l=i s \sqrt{\frac{\rho}{E}} = i \frac{s}{l_a} \sqrt{\frac{m_a}{k_a}}
\end{equation}
The dynamic stiffness matrix of the spectral rod element of length $L$ can be formulated in the same manner as given in the references \cite{GKrishnanBook, DoyleBook}.

\begin{figure}[!h]
\centering
\includegraphics[scale=0.5]{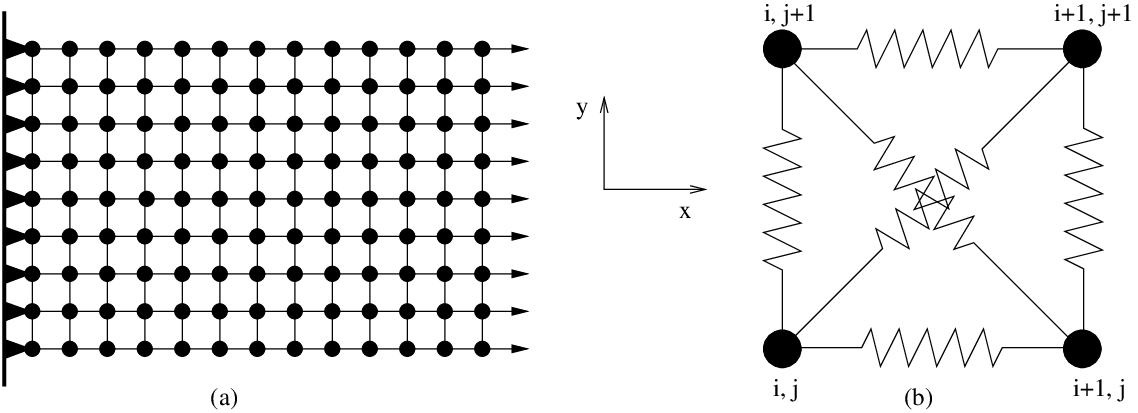}
\caption{Schematic of the $2$D square Bravais lattice system and unit cell.}
\label{atomic2Dlattice}
\end{figure}

\subsection{Spectral Element Formulation for $2$D Atomic Lattices}
Next, a simple $2$D square Bravais lattice is considered (Fig. \ref{atomic2Dlattice}) for modeling wave propagation in two dimensional atomistic system. Continuum approximation for 2D hexagonal close-packed (hcp) lattice can be obtained in a similar manner. The atoms in the $2$D lattice are indexed by $(i,j)\in \mathbb{Z}^2$. The position of the $(i,j)^{th}$ atom in the X-Y plane at time $t$ is given by (Fig. \ref{atomic2Dlattice})
\begin{align}\label{SEBE101}
\left(
  \begin{array}{c}
    x \\
    y \\
  \end{array}
\right)_{i,j}= \left(
                \begin{array}{c}
                ir \\
                jr \\
                \end{array}
                \right) + \left(
                            \begin{array}{c}
                            u(t) \\
                            v(t) \\
                            \end{array}
                            \right)_{i,j} \in \mathbb{R}^2
\end{align}
 Here $r$ is the equilibrium atomic separation and $(u(t), v(t))^T = d_{i,j}(t)$ is out-of-equilibrium atomic displacements along the respective directions in the X-Y plane. Considering nearest-neighbor and diagonal interactions (Fig. \ref{atomic2Dlattice}), the Hamiltonian for the system can be written as \cite{GFriesecke}
\begin{align}\label{SEBE102}
H = \displaystyle \sum_{i,j} (\frac{1}{2m_a}|p_{i,j}|^2 + U_1(|re_1 + d_{i+1,j}- d_{i,j}|) + U_1(|re_2 + d_{i,j+1}- d_{i,j}|) \nonumber\\
 + U_2(|r(e_1+e_2) + d_{i+1,j+1}- d_{i,j}|) + U_2(|r(e_1-e_2) + d_{i+1,j}- d_{i,j+1}|)
 )
\end{align}
Here $|.|$ denotes the Euclidian norm on $\mathbb{R}^2$, $p_{i,j}$ denotes the momentum vector, $e_1$ and $e_2$ are the lattice basis vectors $(1,0)$ and $(0,1)$. The potential $U_1$ corresponds to horizontal and vertical interactions, and $U_2$ is the potential for the diagonal interactions. Here $U_1$, $U_2$ can be any arbitrarily differentiable potential but for continuum approximations we consider equivalent harmonic potential of the form
\begin{align}\label{SEBE103}
U_1(|z|) = \frac{k_1}{2}(|z| - a_1)^2 \nonumber\\
U_2(|z|) = \frac{k_2}{2}(|z| - a_2)^2
\end{align}
where $k_1$ and $k_2$ are spring constants, respectively and $|z|$ denotes the distance of separation between any two nearest neighbor atoms. For any other nonlinear potentials, the curvature of the potential gives the constants of equivalent harmonic springs. The equation of motion of the $(i,j)^{th}$ atom can be written as
\begin{align}\label{SEBE103}
m_a \ddot{d}_{i,j} = -\{-f_1(re_1 + d_{i+1,j} - d_{i,j}) + f_1(re_1 + d_{i,j} - d_{i-1,j}) \nonumber\\
-f_1(re_2 + d_{i,j+1} - d_{i,j}) + f_1(re_2 + d_{i,j} - d_{i,j-1}) \nonumber\\
-f_2(r(e_1+e_2) + d_{i+1,j+1} - d_{i,j}) + f_2(r(e_1+e_2) + d_{i,j} - d_{i-1,j-1}) \nonumber\\
-f_2(r(e_1-e_2) + d_{i+1,j-1} - d_{i,j}) + f_2(r(e_1-e_2) + d_{i,j} - d_{i-1,j+1})
\}
\end{align}
Where the forcing terms are defined as $f_1(z)={U_1}^{\prime}(|z|) \frac{z}{|z|}$ and $f_2(z)={U_2}^{\prime}(|z|) \frac{z}{|z|}$. Considering the $2$D atomistic system as a deep axial waveguide, where the wave is propagating in the X-direction as considered in Friesecke and Matthies \cite{GFriesecke}, we can assume a traveling wave solution of the form
\begin{align}\label{SEBE105}
u(x,y,t)=u_0(x,t), \; \; \;\; \; v(x,y,t)=y\psi_0 (x,t)
\end{align}
Therefore, using Hamilton's principle we have two coupled equations of motion as \cite{DoyleBook, GKrishnanBook}
\begin{align}\label{SEBE10}
C_{11}A \frac{\partial^2 u_0}{\partial x^2} + C_{12}A \frac{\partial \psi_0}{\partial x} = \rho A \frac{\partial^2 u_0}{\partial t^2}, \nonumber\\
C_{66}I \gamma_1 \frac{\partial^2 \psi_0}{\partial x^2} - C_{11}A \psi_0 - C_{12}A \frac{\partial u_0}{\partial x} = \rho I \gamma_2 \frac{\partial^2 \psi_0}{\partial t^2}
\end{align}
Here, $I$, $\rho$, and $A$ are the second moment of area of cross-section, the mass density, and the area of cross-section, respectively. In (\ref{SEBE10}) $\gamma_1$ and $\gamma_2$ are adjustable parameters \cite{DoyleBook, SGopal} and $C_{11}$, $C_{12}$, and $C_{66}$ are stiffness parameters, respectively. For generality, we find the continuum parameters $C_{11}$, $C_{12}$ using following relation as
\begin{align}\label{SEBE11}
 C_{11} = \frac{1}{A_a h}\frac{\partial^2 U_a}{\partial {\epsilon^2}_{xx}}|_{\epsilon_{xx}=0} \;\;\;\;\; C_{12} = \frac{1}{A_a h}\frac{\partial^2 U_a}{\partial \epsilon_{yy} \partial \epsilon_{xx}}|_{\epsilon_{xx}=0,\epsilon_{yy}=0}
\end{align}
where $U_a$ is potential energy per atom in a unit cell under uniform tensile deformation, $A_a=r^2$ is the effective surface area occupied by the atom in the X-Y plane, $h$ is the thickness of the $2$D atomistic lamina and $\epsilon_{xx}=(u_{i+1}-u_i)/r$, $\epsilon_{yy}=(v_{j+1}-v_j)/r$ are strains, respectively. From (\ref{SEBE11}), we obtain the parameters for the Bravais lattice as $C_{11}=(k_1 + k_2)/h$, $C_{12}=k_2/h$, and $C_{66}$ can be obtained from $C_{11}$ and $C_{12}$. The associated boundary conditions for Eq. (\ref{SEBE10}) are specified as
\begin{align}\label{SEBE12}
F = C_{11}A \frac{\partial u_0}{\partial x} + C_{12}A \psi_0, \; \; \; Q = C_{66}I \gamma_1  \frac{\partial \psi_0}{\partial x}
\end{align}
Since there are two independent variables $u_0$ and $\psi_0$, we assume the homogeneous solutions in the form
\begin{align}\label{SEBE12}
u_0 = \hat{u}_0 e^{-(i k x - s t)}, \;\;\;\; \psi_0 = \hat{\psi}_0 e^{-(i k x - s t)}
\end{align}
Substituting these solutions into Eq. (\ref{SEBE10}) and after some manipulations we find the homogeneous general solutions for the deep axial wave guide of length $L$ in Laplace domain as
\begin{align}\label{SEBE17}
\hat{u}_0(x,s)=P_1 \bar{A} e^{-ik_1x} + P_2 \bar{B} e^{-ik_2x} - P_1 \bar{C} e^{-ik_1(L-x)} - P_2 \bar{D} e^{-ik_2(L-x)}\nonumber\\
\hat{\psi}_0(x,s)=\bar{A} e^{-ik_1x} + \bar{B} e^{-ik_2x} + \bar{C} e^{-ik_1(L-x)} + \bar{D} e^{-ik_2(L-x)} \;\;\;\;\;\;\;\;\;\;\;\;\;\;\;
\end{align}
where, $\bar{A}$, $\bar{B}$, $\bar{C}$, and $\bar{D}$ are the constants to be determined from the boundary conditions on the element and $P_n$ are the amplitude ratios defined as
\begin{align}\label{SEBE16}
P_n = i\left[\frac{C_{66} I\gamma_1 k_n^2 + C_{11} + s^2 \rho I\gamma_2}{C_{12}k_n}\right] \;\;\;\;\ n=1,2
\end{align}
Now, the nodal displacements are determined from Eq. (\ref{SEBE17}) as
\begin{align}\label{SEBE18}
\hat{u}_1=\hat{u}_0(0,s), \;\; \hat{\psi}_1=\hat{\psi}_0(0,s), \;\; \hat{u}_2=\hat{u}_0(L,s), \;\; \;\;\; \hat{\psi}_2=\hat{\psi}_0(L,s)
\end{align}
Again, the external force components are computed as
\begin{align}\label{SEBE19}
\hat{F}_1 = -\hat{F}(0,s) , \;\; \hat{Q}_1= -\hat{Q}(0,s), \;\; \hat{F}_2 = \hat{F}(L,s) , \;\; \hat{Q}_2= \hat{Q}(L,s)
\end{align}
Thus the relation between the coefficients vector $\{\texttt{C}\}=[\bar{A} \; \bar{B} \; \bar{C} \; \bar{D}]^T$ and the nodal degrees of freedom $\{\hat{U}\}=[\hat{u}_1 \; \hat{\psi}_1 \; \hat{u}_2 \; \hat{\psi}_2]^T$ is obtained using Eqs. (\ref{SEBE17}) and (\ref{SEBE18}) as
\begin{equation}\label{SEBE20}
\{\texttt{C}\}=[\hat{Q}]\{\hat{U}\}
\end{equation}
Here, $[\hat{Q}]$ is a $4\times 4$ matrix. The relation between nodal loads $\{\hat{F}_b\}=[\hat{F}_1 \; \hat{Q}_1 \; \hat{F}_2 \; \hat{Q}_2]^T$ and nodal degrees of freedom $\{\hat{U}\}$ is obtained using Eqs. (\ref{SEBE17}), (\ref{SEBE19}), and (\ref{SEBE20}) as
\begin{equation}\label{SEBE21}
\{\hat{F}\} = [\hat{R}]\{\texttt{C}\} = [\hat{R}][\hat{Q}]\{\hat{U}\} = [\hat{K}_d]\{\hat{U}\}
\end{equation}
where, $[\hat{R}]$ is a $4\times 4$ matrix which relates the constants $\{C\}$  with nodal loads $\{\hat{F}_b\}$ and $[\hat{K}_d]$ is the exact dynamic stiffness matrix of the $2$D axial element. The details of the formulation and explicit form of the $4\times 4$ stiffness matrix can be found in reference \cite{Gopalakrishnan, GKrishnanBook, DoyleBook}.

\section{Numerical Laplace Transform (NLT)}
The Laplace transform is used as an efficient tool in transient analysis for many years. Numerical Laplace transform (NLT) can suppress the non-causal effects due to time aliasing occurring when a continuous spectra is discretized. Weeks \cite{Weeks} introduced numerical inversion of Laplace transforms using Laguerre functions. Crump \cite{Crump} proposed numerical inversion of Laplace transforms using a Fourier series approximation. Then, Wilcox \cite{Wilcox} introduced numerical Laplace transform and inversion to overcome the difficulties encountered in the inversion of Laplace transform. Since then, many researchers have applied NLT in several fields such as electrical transmission lines \cite{Wedepohl, Ramirez, Moreno}, vibration \cite{Blais}, and acoustics \cite{Igawa, Kishor}. Reference \cite{Murthy} gives the complete spectral FEM formulation using NLT and summarizes its several advantages over SFEM. In the present work, NLT is used to model wave propagation problem in finite structures by means of fast Fourier transform (FFT) algorithm as
\begin{align}\label{NLT7}
L\{f(t)\}=F(s)=F(\sigma + iw)=\mathcal{F} \{f(t)e^{-\sigma t}\}\nonumber\\
L^{-1}\{F(s)\}=f(t)= e^{\sigma t}\mathcal{F}^{-1} \{F(\sigma + iw)\}
\end{align}

The value of $\sigma$ should be positive to damp out the time aliasing error which occurs in NLT. Regarding aliasing errors, Wilcox \cite{Wilcox} proposed the following criterion to select the value of $\sigma$:
\begin{equation}\label{NLT9}
\sigma = 2\pi/T_{max}
\end{equation}
Later, Wedepohl \cite{Wedepohl} proposed the following criterion for $\sigma$:
\begin{equation}\label{NLT10}
\sigma = 2\ln(N)/T_{max}
\end{equation}
This shows the relationship of $\sigma$ with the number of sampling points $N$ in the time signal. For a given value of $T_{max}$, the sampling frequency $\omega_s$ is higher for larger value of $N$. Therefore, the truncation of the spectrum is less significant and the value can be set higher. In this paper, the value of $\sigma$ proposed by Wedepohl \cite{Wedepohl} is adopted for all the numerical examples and is found to give superior results.

\section{Numerical Algorithm and Implementation}
In this section, we present the algorithms for implementation of the spectral multiscale method. An algorithm of the proposed multiscale method is given for clarity. For time integration in the MD computations, we use verlet algorithm as described in \cite{TangBSM, TangPSBSM, Frenkel}. Therefore, we do not explain it here again.

Coarse scale computations for the whole domain $\Omega$ is performed in Laplace domain. The following steps are implemented in the coarse scale computations:
\begin{enumerate}
  \item The appropriate time step size $\Delta t$ and number of sampling points of the simulation $N$ are chosen such that $T_{max}=N\Delta t$.
  \item The sampling frequency given by $f_{max}=1/\Delta t$ which has to be greater than twice the maximum frequency content (in Hz) of the input signal. Thus, the maximum circular frequency is $\omega_{max}=\pi f_{max}$.
  \item The entire frequency domain $[-\omega,\omega]$ is discretized as $\Delta \omega = 2 \pi/T_{max}$.
  \item The forcing function $f(t)$ is sampled and transformed by using NLT (Eq. (\ref{NLT7})).
  \item Computations for spectral responses (displacement, velocity and acceleration) are performed using Eq. (\ref{eq9}) for each frequency $\omega$.
  \item Finally these responses are transformed back to the time domain responses by inverse NLT (Eq. (\ref{NLT7})).
\end{enumerate}

The implementation of the new spectral multiscale method (SMM) is adopted in the following sequence in our computations.
\begin{enumerate}
  \item At first, the coarse grid equation is solved for $\{\hat{d}\}$ in $\Omega$ with a desired time step $\Delta t=m\Delta \tau$ for $N$ sampling points. The mean displacement part $\{\bar{\hat{u}}_G\}$ of the ghost point atoms and $\{\bar{\hat{u}}_B\}$ of the inner boundary atoms are computed in spectral domain using Eq. (\ref{eq5}) and stored for all the sampling steps.
  \item The mean displacements in time-domain are obtained by transforming these spectral mean displacements using inverse NLT.
  \item The MD computation in $\Omega_F$ is run for each time step $\Delta \tau$ in the following manner. At first, MD displacements and mean displacements of interfacial atoms are updated for $i^{th}$ step using information from $(i-1)^{th}$ step. Fine fluctuation part $\{\breve{u}_G(t)\}$ is computed using Eq. (\ref{GLE14}) for the present step. This $\{\breve{u}_G(t)\}$ is added to $\{\bar{u}_G(t)\}$ to get complete $\{u_G(t)\}$, which is then used as the boundary condition to compute the present step acceleration for external forcing in $\Omega_F$. These computation steps are repeated for the next time steps.
\end{enumerate}

\section{Numerical Experiments and Results}
We now present some numerical results to illustrate the efficiency of our proposed spectral multiscale method (SMM). We consider a 1D harmonic lattice (section $6.1$) of finite length, also used by Liu et al. \cite{WagnerBSM, TangBSM, TangPSBSM}, for examining the capability of the new SMM in coupling MD simulation with NLSFEM (continuum) simulation. In section $6.2$, we present a $1$D nonlinear lattice with Lennard-Jones potential and next nearest neighbor interaction to investigate the efficiency of the SMM in bridging simulations with initial displacement condition. We apply the SMM to model wave propagation in a $2$D square Bravais lattice (section $6.3$). In the final example (section $6.4$), wave propagation in a $2$D hexagonal close-packed (hcp) system with a microcrack is modeled using SMM framework.

\subsection{Linear example: 1D harmonic lattice }
Consider a harmonic lattice of length $l=0.25$ m in one space dimension shown in Fig. \ref{harmoniclattice}. We assume this length of the lattice includes $1001$ atoms of mass $m_a=1\times 10^{-6}$ kg with an equilibrium interatomic spacing of $l_a=2.5\times 10^{-4}$ m, first of which from the left is fixed. The position of $i^{th}$ atom at rest is $x_i=(i-1)l_a$. The interatomic force between any two atoms can be modeled as a linear spring force $f_{ji}=k_a(u_j-u_i)$ due to harmonic potential assumption; $k_a=5\times 10^8$ N/m is spring stiffness. The longitudinal wave speed in the harmonic lattice is $C_l=5.5902\times 10^3$ m/s.

\begin{figure}[!h]
    \label{fig:subfigures}
    \begin{center}
        \subfigure[]{%
            \label{mdf}
            \includegraphics[width=0.4\textwidth]{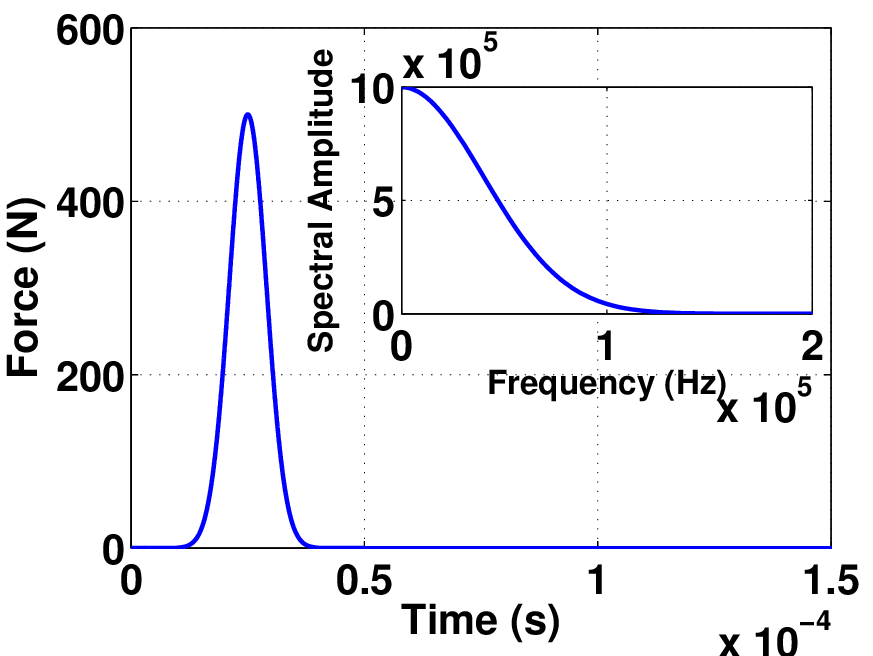}
        }%
        \subfigure[]{%
           \label{thk}
           \includegraphics[width=0.4\textwidth]{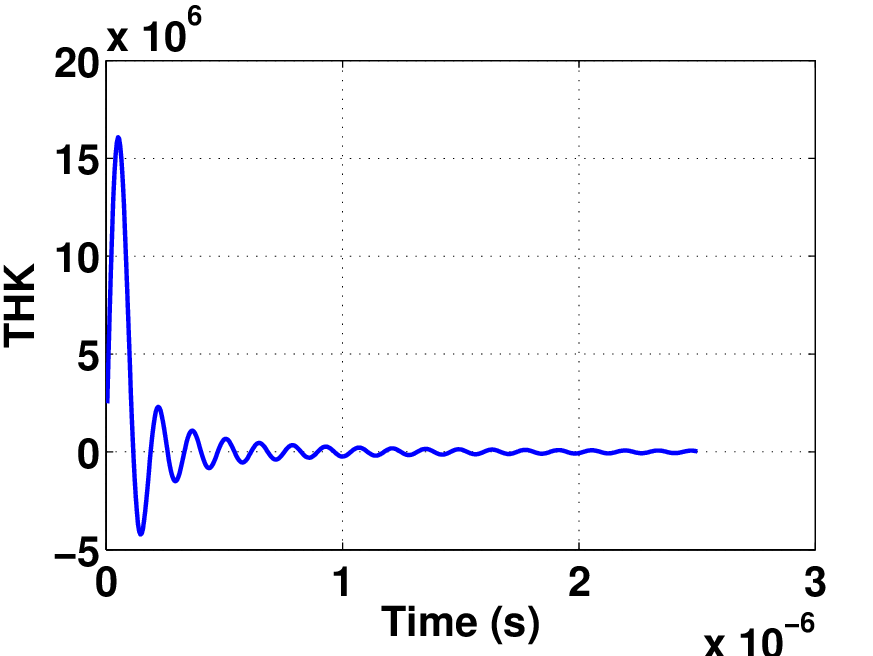}
        }
        \\ 
%
    \end{center}
    \caption{%
        $(a)$ The Gaussian pulse used as input; and $(b)$ the truncated time history kernel $\theta (t)$ used for both the
        harmonic and nonlinear lattice examples.
     }%
     \label{fig12}
\end{figure}

We assume only nearest neighbor interaction for the atomistic system. The interatomic force vector can be written as $ f_{int}(u)=-K_fu$ in matrix format. The stiffness matrix $K_f$ is tri-diagonal. The expression of memory kernel denoted as THK for this $1$D harmonic lattice is derived as
\begin{align}\label{NE14}
\Theta (s) = \frac{1}{2\omega}(2\omega^2 + s^2 - s\sqrt{4\omega^2 + s^2})\nonumber\\
\theta (t) = L^{-1}{\Theta (s)} = \frac{2J_2(2\omega t)}{t}
\end{align}
where $\omega=\sqrt{\frac{k_a}{m_a}}$ and $J_2$ is the second-order Bessel function. We use a truncated time history to reduce the computing load as shown in Fig.\ref{thk}. A Gaussian impulse (Fig. \ref{mdf}) is applied axially on the right most atom. We model the Gaussian force $f_{ext}(t)$ in all the numerical simulations as
\begin{equation}\label{NE13}
f_{ext}(t) = \frac{a_0}{\sigma\sqrt{2\pi}} e^{\frac{1}{2}(\frac{t-\mu}{\sigma})^2}
\end{equation}
In this example, we take $a_0 = 50$, $\sigma = 3.989\times10^{-6}$, and $\mu=25\times10^{-6}$, respectively.

For the multiscale simulation, we choose $\Omega_F$ spanning over $0.22525\leq x \leq0.25$; $x$ in meter, measured from the fixed end. This includes $100$ atoms in the $\Omega_F$ domain. The MD simulation is performed for $1.5\times10^{-4}$ s considering $\Delta \tau= 5\times10^{-9}$ s with a total of $N=30000$ time steps, using the explicit verlet algorithm. The whole domain $0\leq x \leq0.25$ m of the harmonic lattice is considered for coarse scale simulation using a single element of NLSFEM and the simulation is run with $\Delta t=10\times\tau$ for total $N=32768$ sampling points. The mean part of ghost point atoms displacements are computed from coarse scale response using Eq. (\ref{eq5}). The fine fluctuation part of the same is computed using Eq. (\ref{GLE14}).
\begin{figure}[!h]
    \label{fig:subfigures}
    \begin{center}
        \subfigure[]{%
            \label{SMMdisp}
            \includegraphics[width=0.4\textwidth]{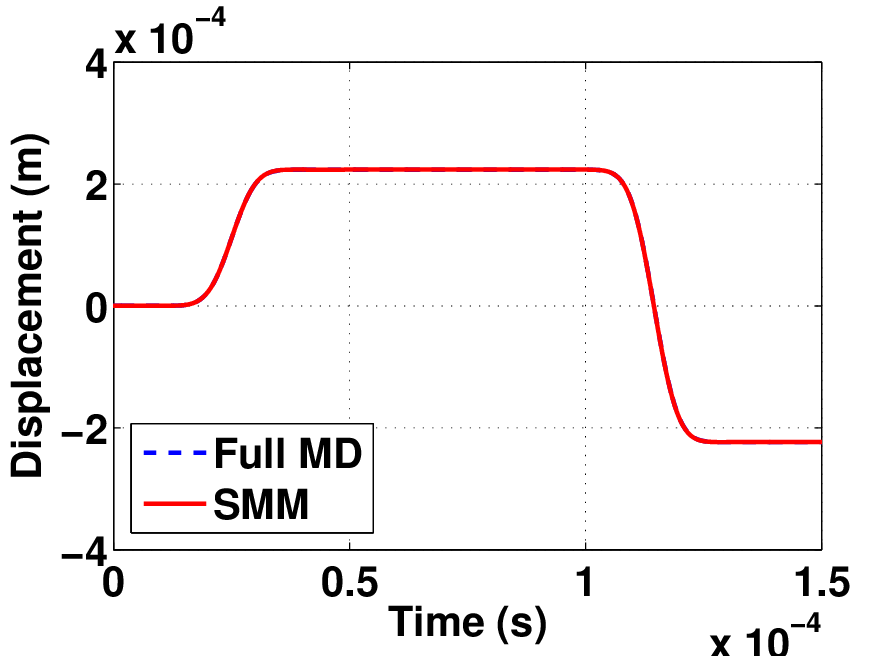}
        }%
        \subfigure[]{%
           \label{SMMvelo}
           \includegraphics[width=0.4\textwidth]{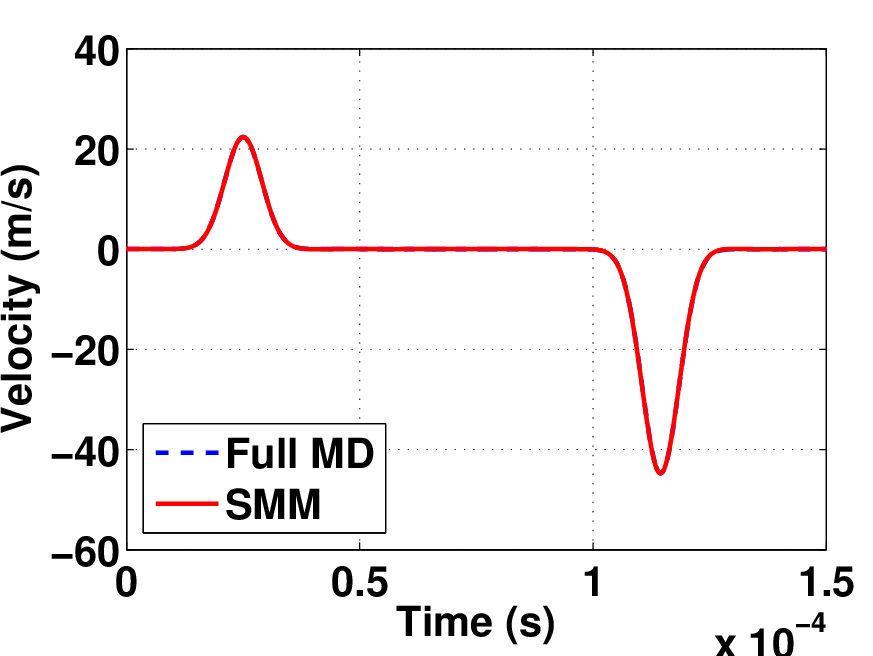}
        }
        \\ 
        \subfigure[]{%
            \label{ffluc}
            \includegraphics[width=0.4\textwidth]{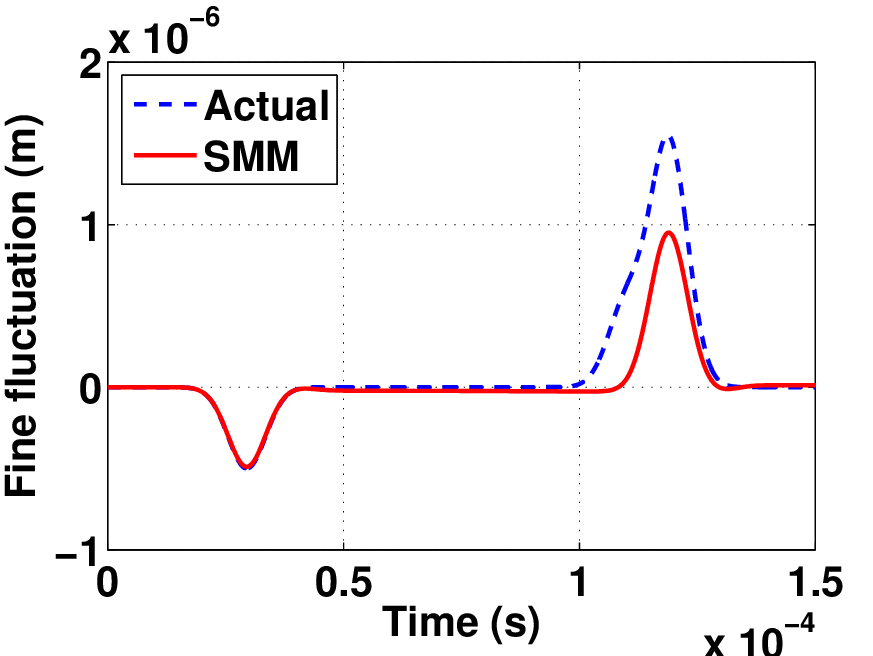}
        }%
        \subfigure[]{%
            \label{HRe}
            \includegraphics[width=0.4\textwidth]{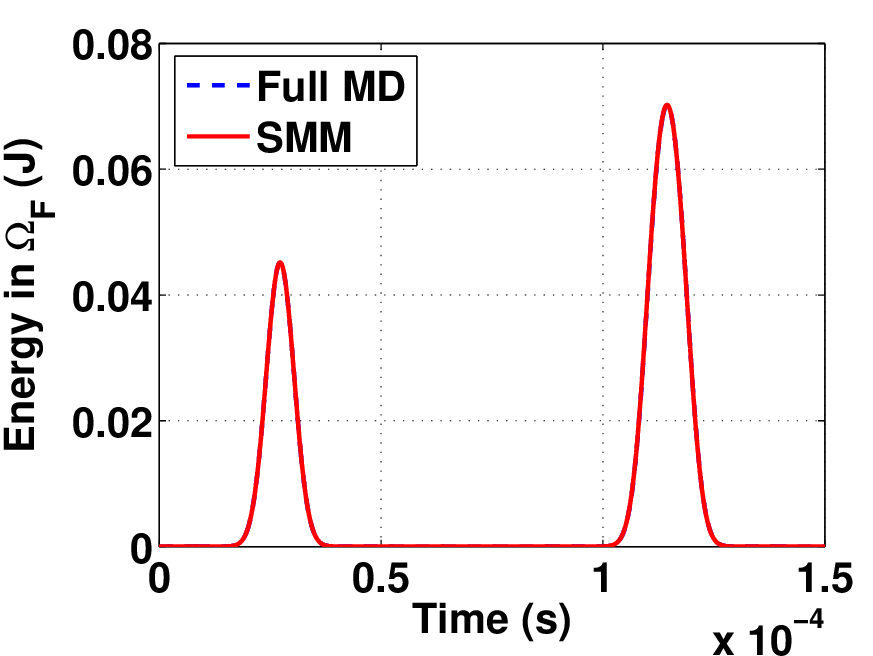}
        }%
    \end{center}
    \caption{%
        The comparison of responses of the free end atom computed by full MD and SMM in the harmonic lattice example: $(a)$ $u(0.25,t)$; $(b)$ $\dot{u}(0.25,t)$; $(c)$ The fine fluctuation of the ghost point atom computed using truncated THK $\theta (t)$; $(d)$ Total energy plot in $\Omega_F$, as a function of time.
       }%
     \label{fig13}
\end{figure}

The full MD simulation gives the standard solution for this case. The analysis results are shown in Fig. \ref{fig13}. The results show the excellent capability of the new spectral multiscale method in simulating the fine scale response of any sub-domain of the harmonic lattice. Both the displacement and velocity responses computed by SMM match accurately with the exact MD responses (Figs. \ref{SMMdisp}-\ref{SMMvelo}). Fine fluctuation of the ghost point atom computed by SMM using truncated THK $\theta (t)$ (Fig. \ref{ffluc}) matches nicely with its actual value, while the wave is propagating towards continuum region from $\Omega_F$. The mismatch in the latter part (Fig. \ref{ffluc}) is due to the error incurred for using truncated THK. The total energy is the sum of the kinetic energy of all the atoms in the MD region, and the potential energy of all the bonds between atoms, plus half of the potential energy in each of the bonds that connects the interfacial atoms with the ghost point atoms \cite{WagnerBSM}. The energy plot in Fig. \ref{HRe} shows the time dependance of the total energy of the MD sub-domain $\Omega_F$. This clearly shows accurate transfer of energy through the MD region and no spurious wave reflection \cite{WE2002, Xli2005, WE2007, Xli2010} occurs at the interface in the numerical experiment. Results of this example show the nice capability of the new SMM in minimizing interfacial reflection. Neglecting the fine fluctuation part in our computation does not have much influence on the fine scale response computed by SMM because the coarse scale model in this case gives the exact solution. If the width of the input Gaussian pulse decreases, then the coarse scale time step  $\Delta t$ should be selected in such a way that it can take account of the frequency content of the pulse. However, the multiscale result does not get affected because of proper selection of $\Delta t$ in the coarse scale computation.

\begin{figure}[!h]
    \label{fig:subfigures}
    \begin{center}
        \subfigure[]{%
            \label{LJ1m}
            \includegraphics[width=0.4\textwidth]{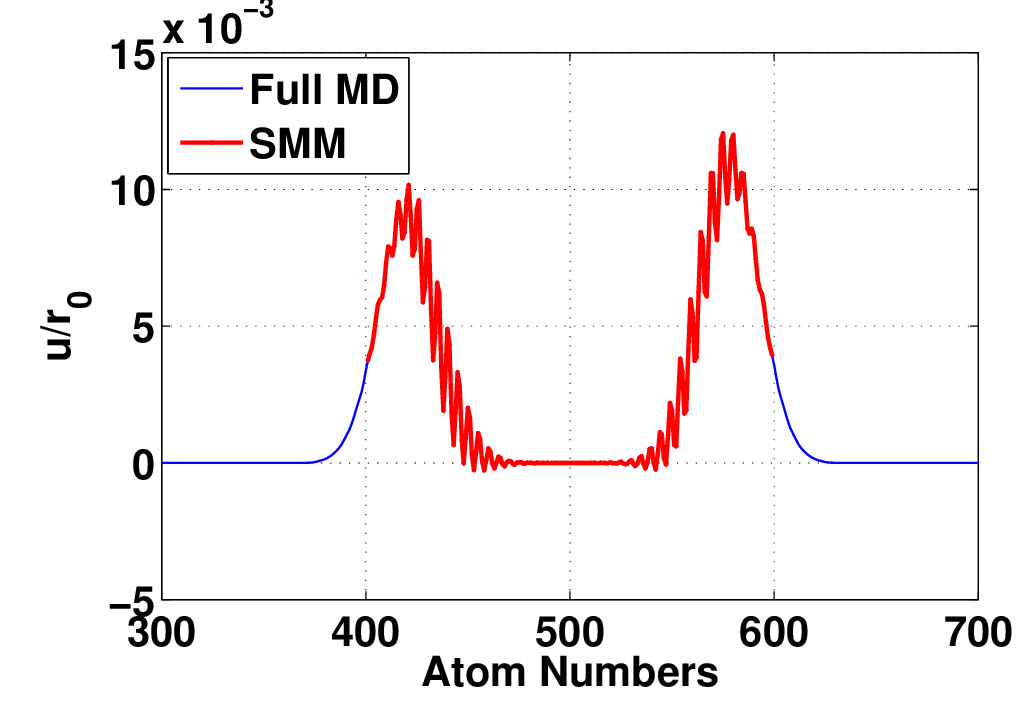}
        }%
        \subfigure[]{%
           \label{LJ2m}
           \includegraphics[width=0.4\textwidth]{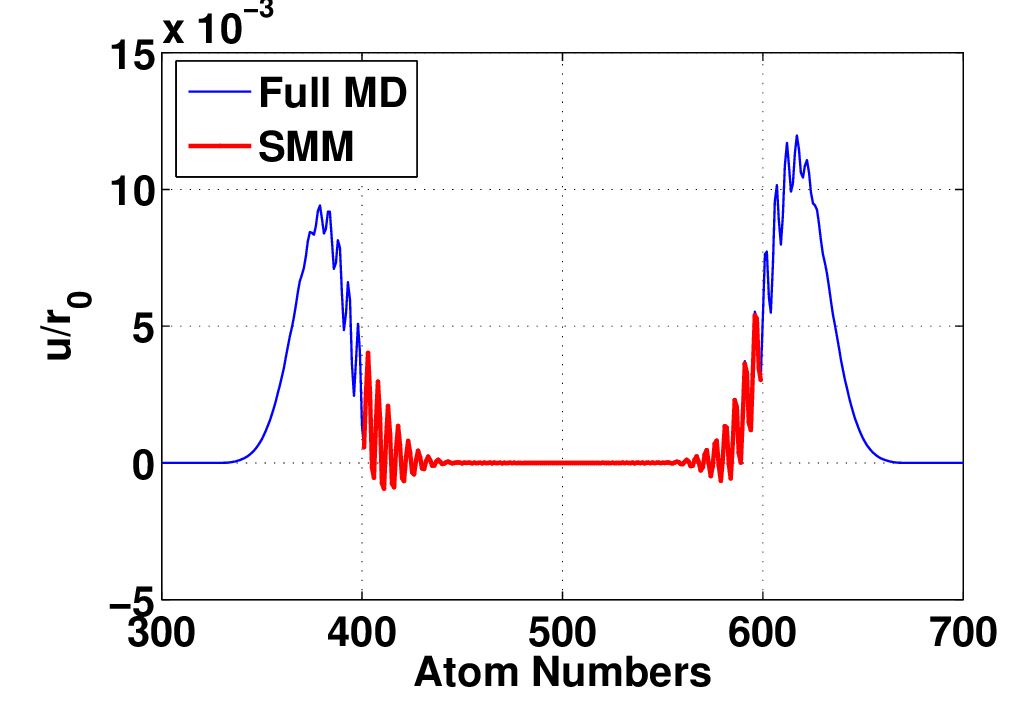}
        }
        \\ 
        \subfigure[]{%
            \label{LJ3m}
            \includegraphics[width=0.4\textwidth]{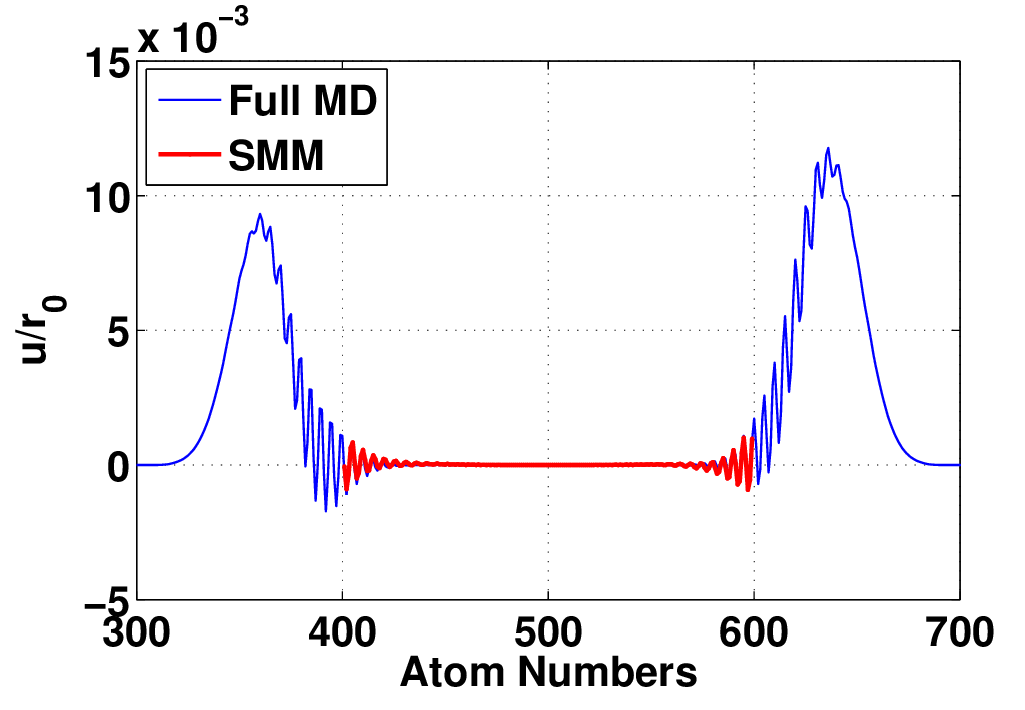}
        }%
        \subfigure[]{%
            \label{LJ4m}
            \includegraphics[width=0.4\textwidth]{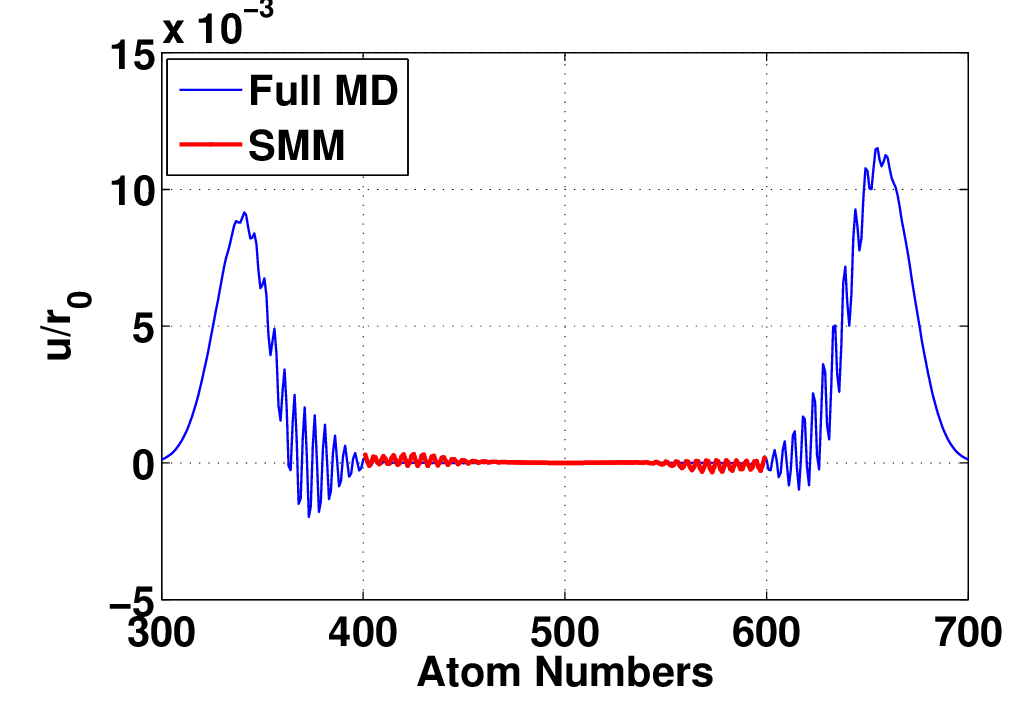}
        }%
    \end{center}
    \caption{%
        Lattice profile of the Lennard-Jones example for $\alpha = 5 \times 10^{-6}$: $(a)$ $\frac{u}{r_0}$ at $t=4\times 10^{-6}$ s;
        $(b)$ $\frac{u}{r_0}$ at $t=6\times 10^{-6}$ s; $(c)$ $\frac{u}{r_0}$ at $t=7\times 10^{-6}$ s; $(d)$ $\frac{u}{r_0}$ at $t=8\times 10^{-6}$ s.
       }%
     \label{LJmdrat}
\end{figure}
\subsection{Nonlinear example $1$: $1$D lattice with Lennard-Jones potential and next nearest neighbor interaction}
We now consider the same $1$D lattice but with a Lennard-Jones potential and displacement initial condition. In this experiment, next nearest neighbor interaction is considered to address the key nonlocal features of atomistic problems. For any initial displacement $\{u\}=(u_1,...,u_n)^T$, the potential function is
\begin{align}\label{NE5}
U(u) = 4\varepsilon \displaystyle \sum_{n} \left[\left(\frac{\sigma}{r_0+u_{n+1}-u_n}\right)^{12} - \left(\frac{\sigma}{r_0+u_{n+1}-u_n}\right)^{6} \right]
\end{align}
Here $r_0$ is the equilibrium atomic separation, $\sigma$ is the collision diameter, and $\varepsilon$ is the bonding energy. In this example, $r_0=2.5\times10^{-4}$ m, $\sigma=r_0/2^{\frac{1}{6}}$ m, and $\varepsilon = 0.35$ J. We take $\Omega_F =[401r_0,599r_0]$ with $199$ atoms inside. We consider the initial wave (displacements) lies in the $\Omega_F$ region. Therefore, we take $199$ coarse grid points in $\Omega_F$ to capture the initial wave energy properly. For next nearest neighbor consideration, we place $6$ more grid points in $\Omega_C$, comprising a total of $205$ nodes in the entire domain $\Omega$. We take $2$ NLSFEM elements spanning over $0 \leq x \leq 399r_0$ and $601r_0 \leq x \leq 0.25$, respectively, and $202$ more conventional FEM elements spanning over $399r_0 \leq x \leq 601r_0$ to capture the initial wave energy properly in our coarse scale computation.
\begin{figure}[!h]
    \label{fig:subfigures}
    \begin{center}
        \subfigure[]{%
            \label{LJ1m}
            \includegraphics[width=0.4\textwidth]{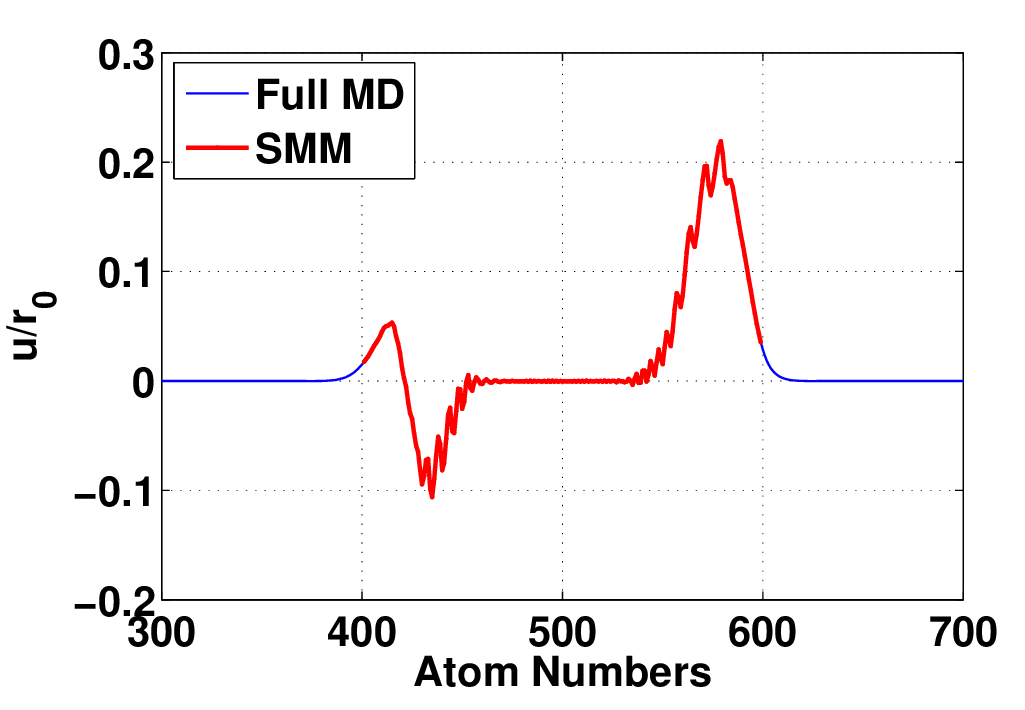}
        }%
        \subfigure[]{%
           \label{LJ2m}
           \includegraphics[width=0.4\textwidth]{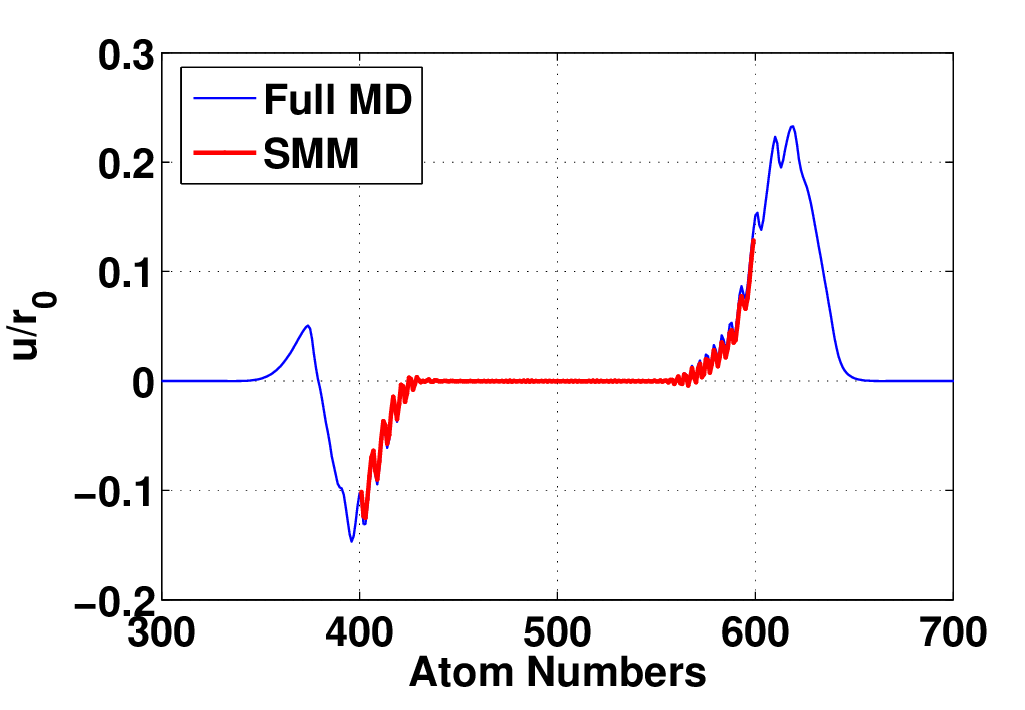}
        }
        \\ 
        \subfigure[]{%
            \label{LJ3m}
            \includegraphics[width=0.4\textwidth]{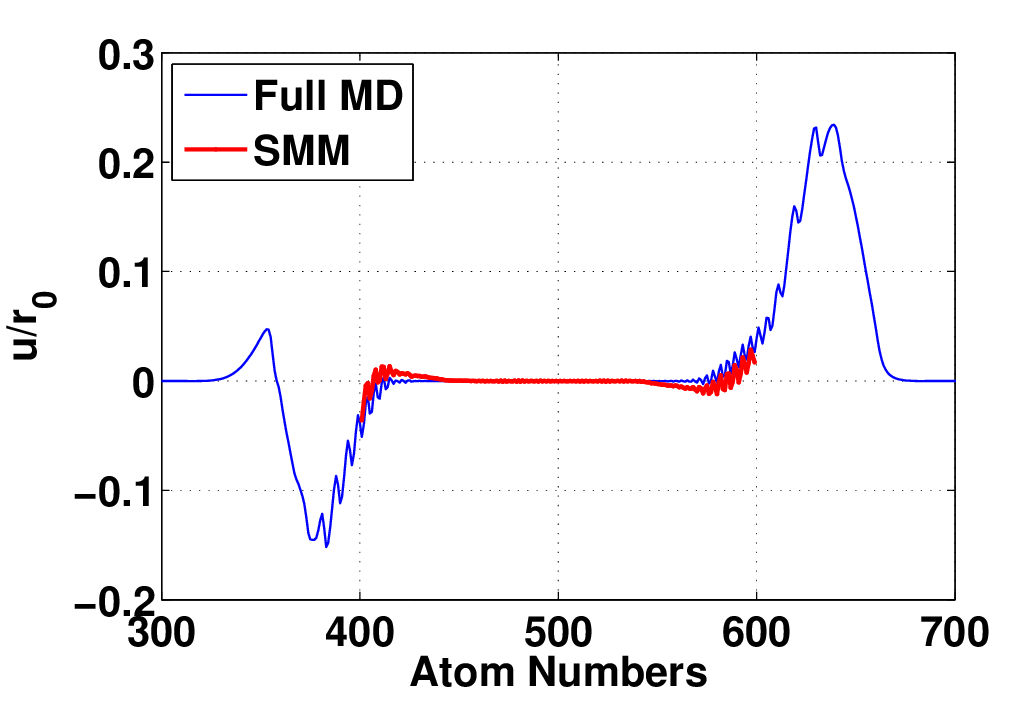}
        }%
        \subfigure[]{%
            \label{LJ4m}
            \includegraphics[width=0.4\textwidth]{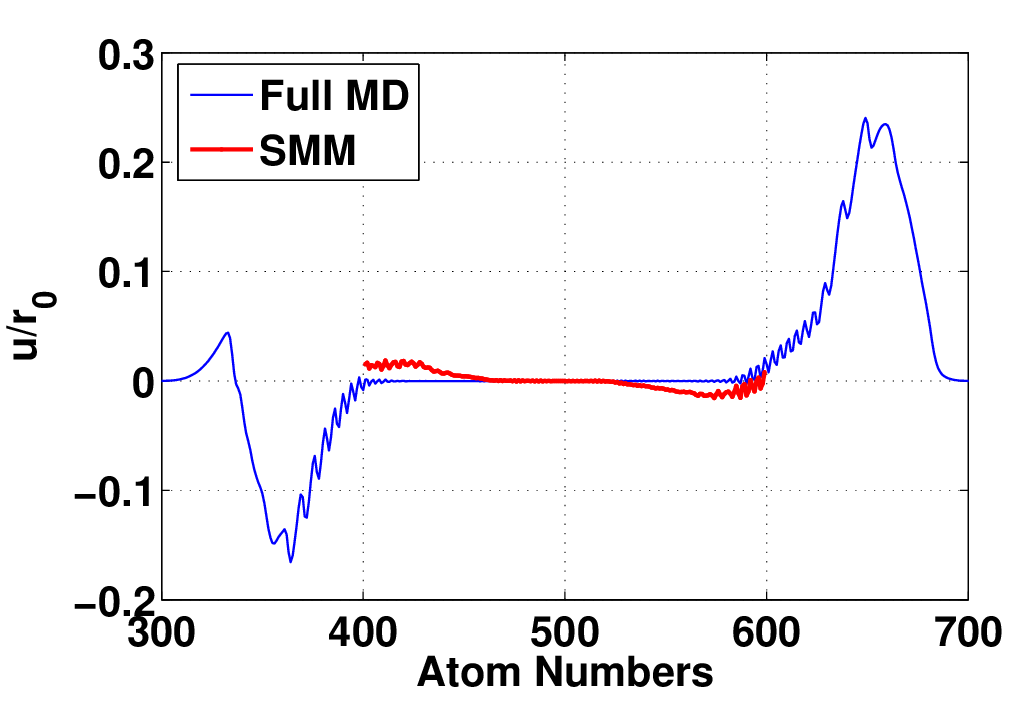}
        }%
    \end{center}
    \caption{%
        Lattice profile of the Lennard-Jones example for $\alpha = 5 \times 10^{-5}$: $(a)$ $\frac{u}{r_0}$ at $t=4\times 10^{-6}$ s;
        $(b)$ $\frac{u}{r_0}$ at $t=6\times 10^{-6}$ s; $(c)$ $\frac{u}{r_0}$ at $t=7\times 10^{-6}$ s; $(d)$ $\frac{u}{r_0}$ at $t=8\times 10^{-6}$ s.
       }%
     \label{LJstrg}
\end{figure}

We consider the initial displacement condition as
\[ u_n = \left\{
  \begin{array}{l l}
    \alpha \frac{e^{-(250n_x)^2}-e^{-6.25}}{1-e^{-6.25}}(1+0.2cos(1800\pi n_x)) & \quad \text{if $401 \leq n\leq 599$}\\
    0 & \quad \text{elsewhere}
  \end{array} \right.\]
where $u_n$ is the displacement of $n^{th}$ atom and $n_x=(n-500)r_0$. In this experiment, we use two different values of $\alpha$ to control the strength of nonlinearity. We take $\alpha = 5 \times 10^{-6}$ for a moderate nonlinearity and $\alpha = 5 \times 10^{-5}$ for strong nonlinearity. The curvature of $U(u)$ at $r_0$ gives $k_a$ in this case. We use the multiscale interfacial condition derived for harmonic lattice, which is not the exact condition required for this case. We compute the fine fluctuation part of the first ghost neighbor atoms through the time history convolution of fine fluctuations of interfacial atoms. Fine fluctuation part of the next neighbor atoms is computed through the time history convolution of fine fluctuations of the first ghost atoms. The MD simulation is performed considering $\Delta \tau=2.5 \times 10^{-9}$ s for total $12000$ steps. We take coarse scale time step length $\Delta t=10 \times \tau$ in the present experiment. Results (Figs. \ref{LJmdrat}-\ref{LJstrg}) agree with full MD solutions quite satisfactorily. For moderate nonlinear case, SMM results (Fig. \ref{LJmdrat}) agree with the exact solution very well. In the strong nonlinear case, the maximum bond strain exceeds $6\%$. The Fig. \ref{LJstrg} shows considerably good agreement between SMM and MD results. On the basis of strain rate and accuracy, results of this test is comparable to results of Tang \cite {STangBC} and Li \cite {Xli2010}. Although small reflection occurs at the interface due to the failure of inexact interfacial condition, the SMM is still capable of reproducing the responses in $\Omega_F$ region satisfactorily.

\subsection{Nonlinear example $2$: $2$D Bravais lattice}
In this example, we apply SMM in $2$D atomistic system. We take a $2$D square Bravais lattice (Fig. \ref{atomic2Dlattice}) with $500 \times 25$ atoms of mass $m_a = 1 \times 10^{-8}$ kg lying in X and Y directions, respectively. Although we take harmonic potentials $V_1$ and $V_2$, the Hamiltonian equations of motion are still nonlinear due to the frame-indifference of the interatomic forces \cite{GFriesecke}. In this test, we assume equilibrium atomic separation $r = 2 \times 10^{-4}$ m, spring constants $k_1 = 5\times 10^6$ N/m and $k_2 = 1\times 10^6$ N/m, thickness $h = 1\times 10^{-5}$. The adjustable parameters assumed as $\gamma_1 = 1.2$ and $\gamma_2 = 1.75$ \cite{SGopal}.
\begin{figure}[!h]
    \label{fig:subfigures}
    \begin{center}
        \subfigure[]{%
            \label{br1}
            \includegraphics[width=0.4\textwidth]{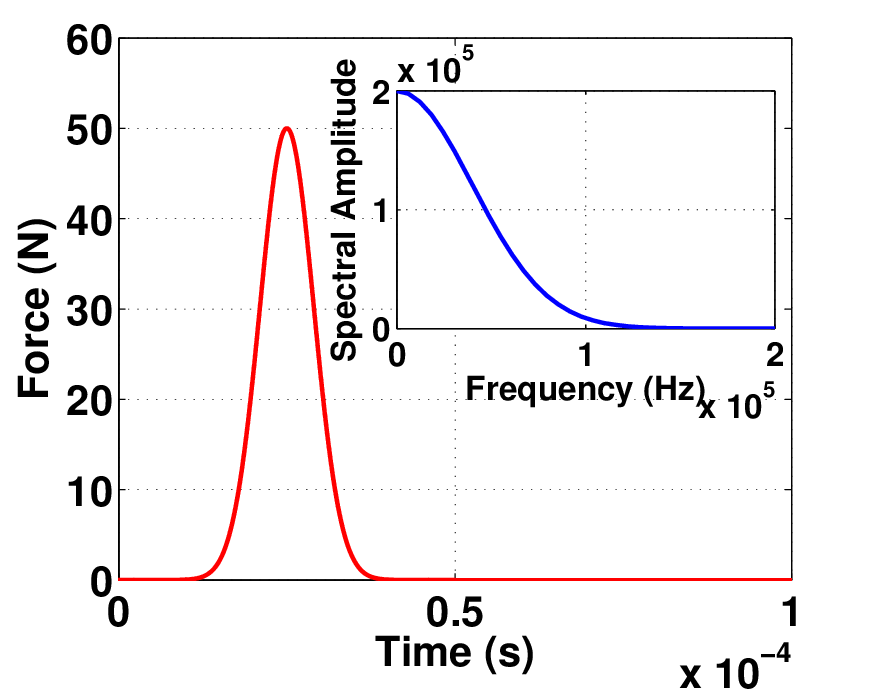}
        }%
        \subfigure[]{%
           \label{br2}
           \includegraphics[width=0.4\textwidth]{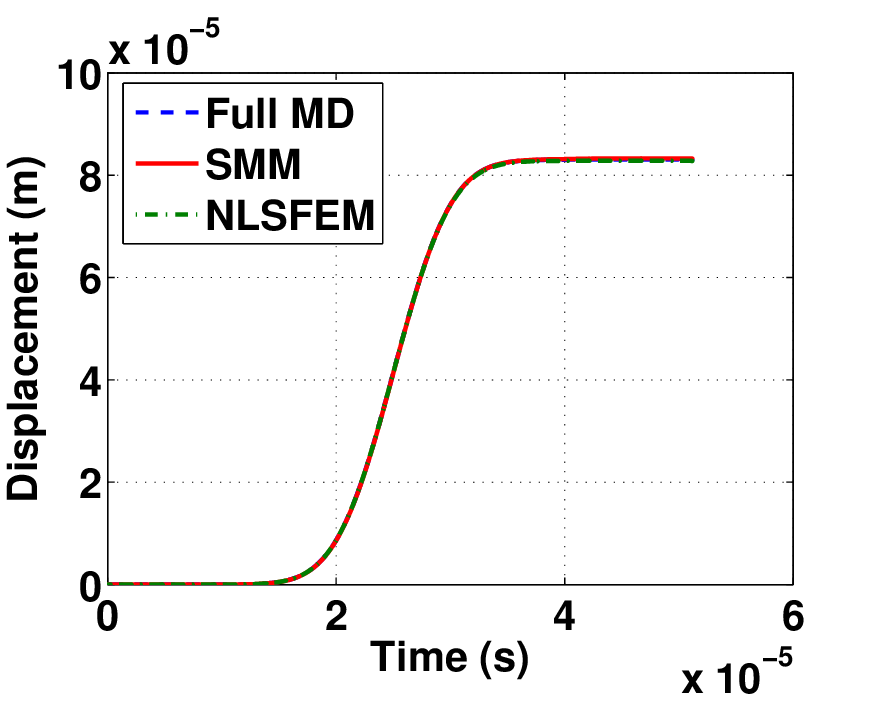}
        }
        \\ 
        \subfigure[]{%
            \label{br3}
            \includegraphics[width=0.4\textwidth]{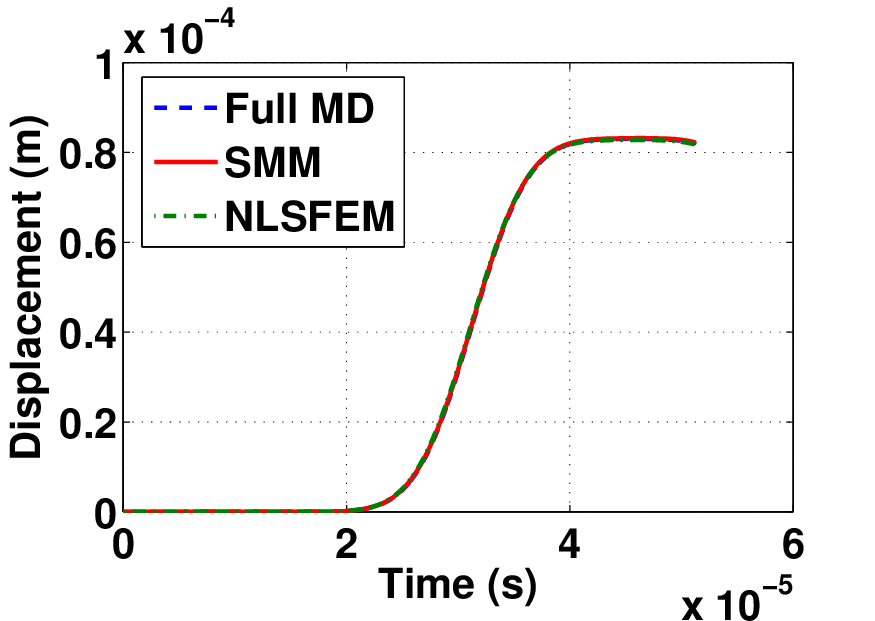}
        }%
        \subfigure[]{%
            \label{br4}
            \includegraphics[width=0.4\textwidth]{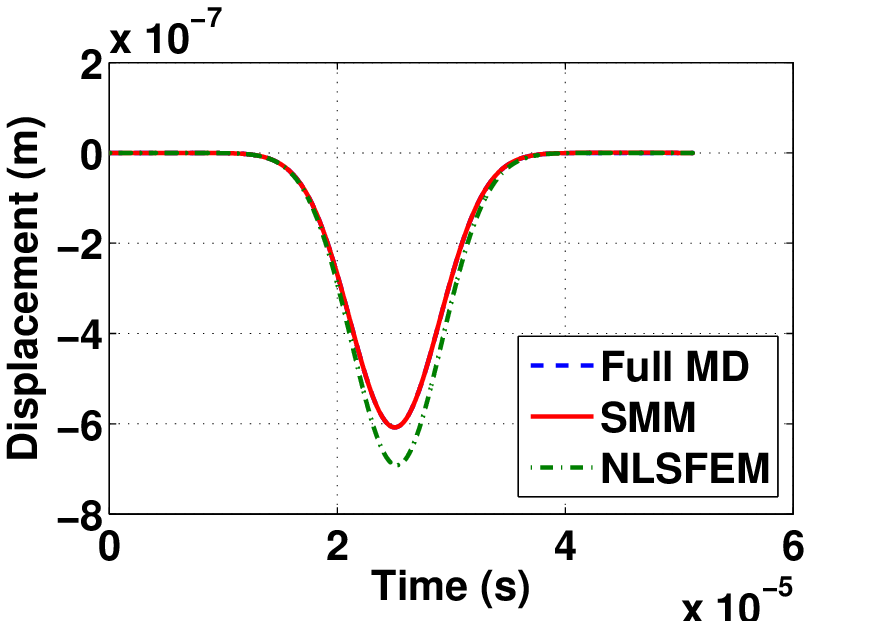}
        }%
    \end{center}
    \caption{%
        The comparison of full MD and SMM responses in $2$D Bravais lattice example: $(a)$ The forcing used in $2$D lattice example;
        $(b)$ Displacement of the top rightmost atom in X direction $u_{500,25}(t)$;
        $(c)$ Displacement of the left boundary atom in X direction $u_{350,13}(t)$;
        $(d)$ Displacement of the top rightmost atom in Y direction $v_{500,25}(t)$.
       }%
     \label{Bravais}
\end{figure}

Here, only the region spanning over $350 \leq i \leq 500$ and $1 \leq j\leq 25$ is considered as $\Omega_F$, where $i$ and $j$ represent the atom's location in the X-Y plane as shown in Fig. \ref{atomic2Dlattice}. The whole domain ($\Omega$) is considered for coarse scale computations and only one NLSFEM element is taken for the entire domain ($\Omega$). Each atom on the right side (Fig. \ref{atomic2Dlattice}) is pulled by an external forcing function given as
\begin{equation}\label{NE31}
f_{ext}(t) = \frac{1}{n_{ac}}\frac{a_0}{\sigma\sqrt{2\pi}} e^{\frac{1}{2}(\frac{t-\mu}{\sigma})^2}
\end{equation}
Here $n_{ac} = 25$ is the total number of atoms lying in a column along the Y direction. In this example, the values $a_0 = 5 \times 10^{-3}$, $\sigma = 3.989\times 10^{-6}$, and $\mu = 25 \times 10^{-6}$ are assumed, respectively. We compute the entire domain for total $T = 5.12 \times 10^{-5}$ s in $N = 20480$ time steps with $\Delta t = 2.5 \times 10^{-9}$ s.
\begin{figure}[!h]
    \label{fig:subfigures}
    \begin{center}
        \subfigure[]{%
            \label{br21}
            \includegraphics[width=0.4\textwidth]{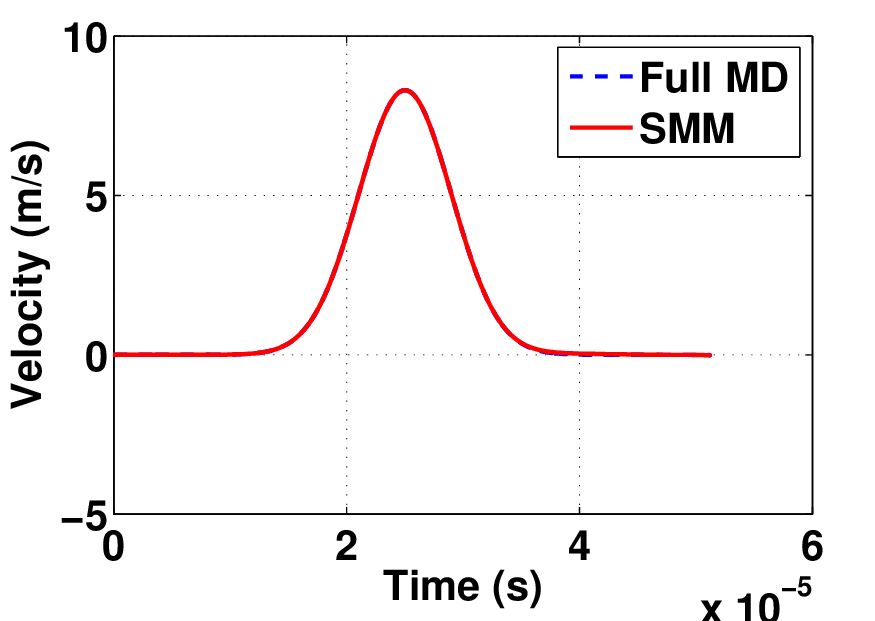}
        }%
        \subfigure[]{%
           \label{br22}
           \includegraphics[width=0.4\textwidth]{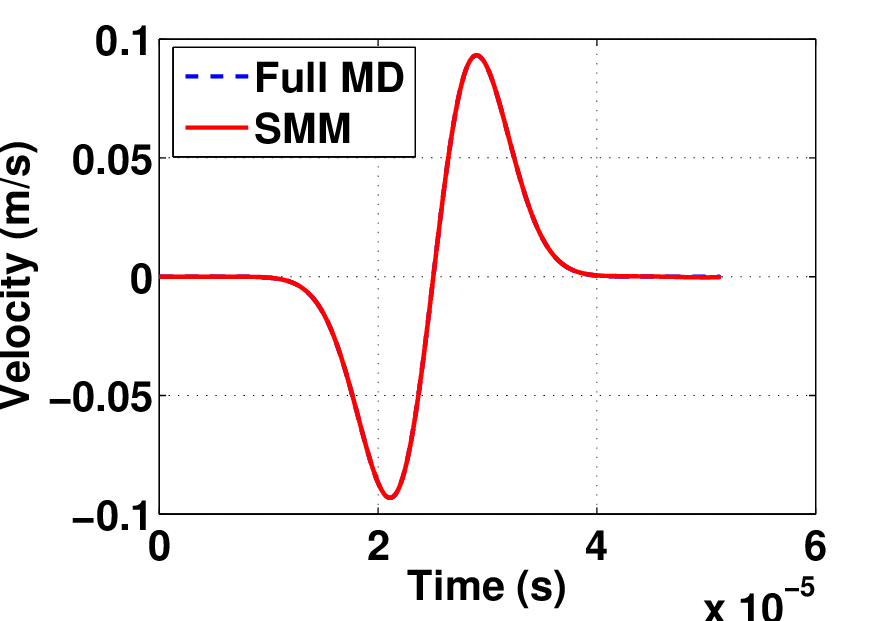}
        }
        \\ 
%
    \end{center}
    \caption{%
        The comparison of full MD and SMM responses in $2$D Bravais lattice example:
         $(a)$ Velocity of the top rightmost atom in X direction $\dot{u}_{500,25}(t)$;
         $(b)$ Velocity of the top rightmost atom in Y direction $\dot{v}_{500,25}(t)$.
     }%
     \label{brv2}
\end{figure}

The focus here is to use simple interfacial condition to avoid computational load. Observing the displacements of $2$D atomistic lamina along X direction are function of $x$, the same interfacial condition developed for $1$D harmonic lattice is again used here for this $2$D lattice system. The displacement of the $(i,j)^{th}$ ghost atom location are assumed to be coupled only
\begin{table}[!h]
\caption{Comparison of computation times in the $2$D Bravais lattice example.}
\centering
\begin{tabular}{|c c c c c|}
  \hline
   Method &  Number of            &  DOF in $\Omega_F$ &  Response  &  Computation  \\
          & atoms ($n_x \times n_y$) &                    & length (s) & time (s)     \\
  \hline
   Full MD &  $500 \times 25$ & $25000$ &  $5.12\times10^{-5}$  &  $4366.4$  \\
   SMM &  $500 \times 25$ & $7500$ &  $5.12\times10^{-5}$ &  $2636.9$  \\
   NLSFEM &  $500 \times 25$ & $6$ &  $5.12\times10^{-5}$ &  $2.26$  \\
  \hline
\end{tabular}
\label{tableBR}
\end{table}
to that of the $(i+1,j)^{th}$ location on right side of the interface. Therefore, we compute the fine fluctuations of $(i,j)^{th}$ ghost atom through the time history convolution of corresponding fluctuations of $(i+1,j)^{th}$ boundary atom. Thus, we use the same memory kernel (Eq. \ref{NE14}), but with a modified $\omega =\sqrt{\frac{k_1(k_1+2k_2)}{m_a(k_1+k_2)}}$ which gives very satisfactory results in this case. Here, $r\omega$ is the axial wave speed in the $2$D lattice. Results (Figs. \ref{Bravais}-\ref{brv2}) show very good agreement of SMM results with the full MD solutions. Here again, only $1$ NLSFEM element is used in our coarse scale computation which significantly reduces the computation time (Table \ref{tableBR}) and gives very good approximation of the mean displacements at ghost atom locations. The Table \ref{tableBR} shows that SMM works nicely for a large system with appreciably less computation cost. In this regards, SMM is an incomparably advantageous method.

\begin{figure}[!h]
\centering
\includegraphics[scale=0.6]{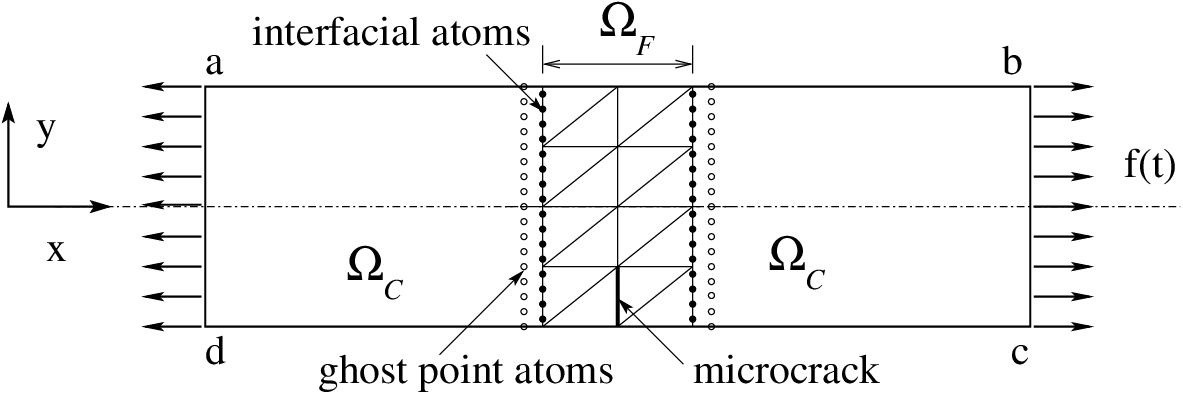}
\caption{Schematic of the $2$D triangular lattice system with microcrack.}
\label{hcp}
\end{figure}
\subsection{Nonlinear example $3$: $2$D Hexagonal close-packed lattice (hcp) with a microcrack}
In this example, we briefly show the capability of SMM in modeling the dynamics of a $2$D equilateral triangular lattice system with a microcrack (Fig. \ref{hcp}). We take a $2$D triangular lattice system with hexagonal nearest neighbor interaction (Fig. \ref{hcp}). The $2$D lamina consists of $20250$ atoms of mass $m_a = 1 \times 10^{-8}$ kg lying in total $500$ columns parallel to the Y directions. Every odd and even numbered columns are having $40$ and $41$ atoms, respectively. The $2$D hcp lamina has a length of $L_x= 8.6429 \times 10^{-2}$ m, a breadth of $L_y= 8.1 \times 10^{-3}$ m and we assume a thickness $h = 1\times 10^{-4}$ m. The lamina has a edge pre-crack, parallel to the close-packed direction, in between the column numbers $250$ and $251$ as shown in Fig. \ref{hcp}. The microcrack has a depth of $10 r$ along Y direction (Fig. \ref{hcp}), where $r = 2 \times 10^{-4}$ m is the equilibrium atomic separation. We assume the crack will not grow during the dynamic simulation. Therefore, we consider a harmonic potentials $U_1$ of spring constant $ k_1 = 6\times 10^6$ N/m. The continuum parameters computed for this hcp system are $C_{11}=\frac {3\sqrt{3}}{4} \frac {k_1}{h}$ and $C_{12}=\frac {\sqrt{3}}{4} \frac {k_1}{h}$. The lamina is pulled axially by Gaussian impulses as shown in Fig. \ref{hcp}. We choose the forcing to be same as the forcing of the previous $2$D example.

 \begin{figure}[!h]
    \label{fig:subfigures}
    \begin{center}
        \subfigure[]{%
            \label{br21}
            \includegraphics[width=0.5\textwidth]{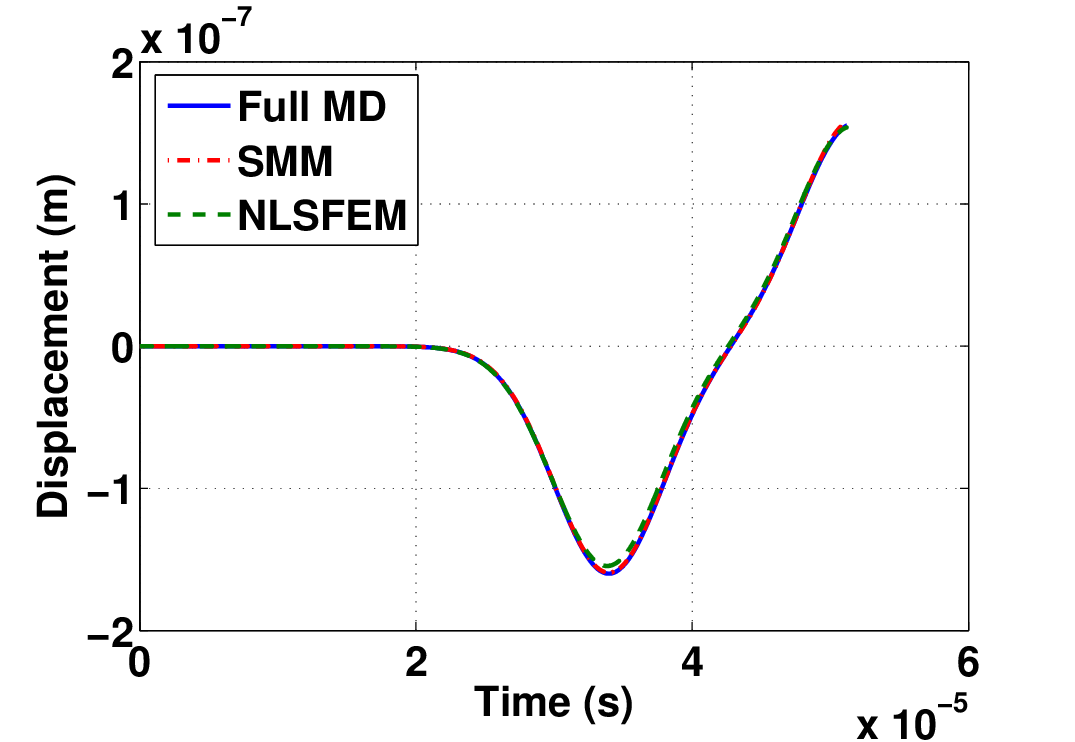}
        }%
        \subfigure[]{%
           \label{br22}
           \includegraphics[width=0.5\textwidth]{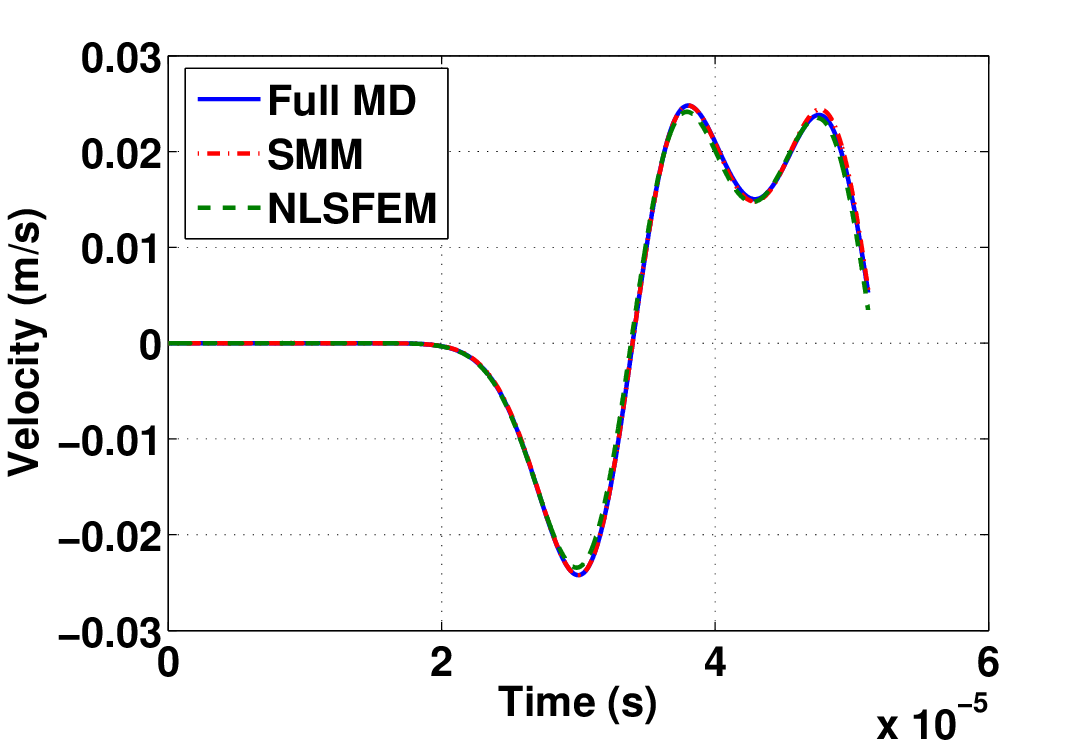}
        }
        \\ 
%
    \end{center}
    \caption{%
        The comparison of full MD and SMM responses in $2$D triangular lattice example:
         $(a)$ Displacement of the top atom in the $250^{th}$ column in X direction $u_{250,41}(t)$;
         $(b)$ Velocity of the top atom in the $250^{th}$ column in X direction $\dot{u}_{250,41}(t)$.
     }%
     \label{hcp2}
\end{figure}
Here, only the region spanning over column number $151 \leq i \leq 351$ is considered as $\Omega_F$ and the whole domain ($\Omega$) for coarse scale computations. The $\Omega_F$ is modeled using $160$ conventional quadratic triangular elements and the $\Omega_C$ on either side of $\Omega_F$, are  modeled using $2$ numbers of $1$D NLSFEM element for coarse scale. The $1$D-$2$D bridging of conventional $2$D FEM and NLSFEM is done adopting a previously developed technique \cite{ANag1,ANag2}. The adjustable parameters assumed as $\gamma_1 = 1.2$ and $\gamma_2 = 1.75$ \cite{SGopal}. The MD simulation is run for $T= 8.192 \times 10^{-5}$ s with $\Delta \tau= 2.5 \times 10^{-9}$ s in total $20480$ steps. The coarse scale simulation is run for $N=20480$ sampling points with $\Delta t= 2.5 \times 10^{-9}$ s. Here also, we use simple interfacial condition as previous example to avoid computational load. We use the same memory kernel (Eq. \ref{NE14}), but with a modified $\omega =\sqrt{\frac{2}{3} \frac{k_1}{m_a}}$. Although this interfacial condition is inexact, still it gives very satisfactory results in this case. Results are depicted in Fig. \ref {hcp2}.

\section{Discussion}
In this paper, a new multiscale method is presented for the atomistic-continuum coupled simulation of a complex system involving different physics at different spatial and/or temporal scales. The novel feature in SMM is its potent continuum model for the coarse scale which is derived directly considering the physics at fine scale to capture the complex phenomena of fine scales. This continuum equation is then, solved semi-analytically in Laplace domain (NLSFEM) to obtain the mean displacement part of the necessary interfacial conditions for the windowed fine scale analysis. In NLSFEM framework the huge number of fine scale degrees of freedom (DOFs) can be reduced to a small number of coarse scale DOFs without hampering the accuracy of the solution. Thus, this method is capable of simulating many realistic systems of any size with localized defects (cracks or dislocations) for a propagating wave of very high frequency with limited computational cost.

To find the time dependent interfacial conditions, the proposed SMM assumes decomposition of the total fine scale spectral (Laplace domain) displacement into a mean part and a fine fluctuation part. The mean part of the response at any crucial location is obtained directly by mapping the coarse scale spectral solution. The remaining fine fluctuation part at ghost nodes/atoms can be obtained by the equation of fine fluctuations (\ref{GLE14}), which involves time history convolution. For nonlinear material systems, we can use the Eq. (\ref{GLE14}) to find the fine fluctuation part at ghost point nodes/atoms. The new spectral decomposition of total displacement field in the proposed SMM makes the effect of fine fluctuation part practically negligible for many real systems.

In modeling discrete systems, the equivalent continuum model for the coarse scale is derived from the governing physics at fine scale. In the current framework, it is difficult to incorporate non-linearity of the fine scale into the coarse scale spectral equation of motion. Therefore, we have made the following approximations for developing the coarse scale equation and interfacial condition:

\begin{itemize}
  \item  Only the linear part of the fine scale equation of motion is incorporated in the spectral form of fine scale equation of motion.
  \item  Only the linear effect of fine fluctuation at the ghost point atoms across the interface on the fine scale sub-domain is taken into account.
\end{itemize}
The possible error sources in our derivation and numerical experiments are as follows:
\begin{itemize}
  \item  The time aliasing error introduced due to analysis in spectral domain.
  \item  The assumption of internal force to be linear in displacement for modeling interfacial interaction.
  \item  The error introduced by the approximation of projection operator $\hat{P}$ to be an identity matrix in the spectral form of the equation of motion for fine fluctuation in the interfacial interaction modeling.
\end{itemize}

Considering the approximations above, the response of a finite 1D harmonic lattice is simulated quite accurately without reflection at the interfaces. With the spectral decomposition of displacement field, the energy interchange between the fine fluctuation and mean displacement across the interface is significantly reduced to a negligible amount. The results of nonlinear $1$D Lennard-Jones lattice with next nearest neighbor interaction shows promising capability of the SMM in capturing nonlocal features of atomistic problems. In linear elasticity the limiting criterion for strain is $1\%$. However, the $1$D Lennard-Jones lattice is simulated beyond $6\%$ bond strain using SMM quite satisfactorily. Results of this test is comparable to results of Tang \cite {STangBC} and Li \cite {Xli2010}. The SMM results of 2D Bravais lattice example show very satisfactory agreement with the full MD results. In this case SMM is computationally very advantageous because a huge number of atoms/nodes of the fine scale model can be included within a single coarse scale element without any deterioration in the simulation results. Again, massive parallelization is viable in NLSFEM framework. Thus, SMM works nicely for a large system with appreciably less computational cost. The SMM results of 2D Hexagonal close-packed system also show very satisfactory agreement with the full MD results.

In NLSFEM framework it is difficult to model 2D and 3D element directly, thus conventional 2D and 3D element should be used with 1D NLSFEM. The time aliasing problem can be handled considering time sampling rate $\Delta t$ to be very small and the number of sampling points $N$ to be very high and with an appropriate value of the constant $\sigma$. These do not significantly add to either computational cost or additional memory requirement.

\section{Conclusions}
\label{Conclusions}
The proposed new multiscale method (SMM) can simulate the dynamics of many practical systems with defects or localized inhomogeneity. The motivating features of the proposed multiscale method are as follows:
\begin{enumerate}
  \item  The SMM formulation is clear and straightforward. It can simulate very high frequency wave propagation with very high accuracy specifically for many systems of crystalline solids with or without crack.
  \item  The proposed method is very robust and has general applicability. In this method, a huge number of atoms/nodes in the fine scale model can be included within a single coarse scale element without any deterioration in the simulation results. Thus, it works nicely for a large system with appreciably less computation cost. In this regard, SMM is an incomparably advantageous method.
  \item  The SMM can run coupled simulations for external input forces as well as for the initial conditions. Any arbitrary positioning of the fine scale domain $\Omega_F$ in the whole domain $\Omega$ does not create any complication in deriving the essential interfacial condition. Moreover, SMM is insensitive to the size of $\Omega_F$.
  \item  The SMM can run coupled simulations of systems with free-free boundary conditions for external input forces.
  \item  No handshaking region is required at all in the modeling of interfacial interactions and the coarse scale continuum model of $\Omega_C$ in spectral domain can be coupled directly with the fine scale model of $\Omega_F$ quite satisfactorily in many computations.
\end{enumerate}

Incorporating the nonlinearity of fine scale equations of motion into the coarse scale spectral equation of motion is one of the main challenges in further developing the SMM. An efficient interfacial condition for nonlinear systems along with a potent coarse scale model can practically solve the snag of multiscale modeling. Another issue is the modeling of finite temperature problems in this SMM framework to take account of the temperature nonuniformity over the whole domain and its effects on the mechanical properties of the system. Modeling phase transformation and solid fracture problems in SMM framework is our next goal. Furthermore, because of its computational efficiency the SMM approach is suitable for stochastic simulations. We shall investigate all these issues in the future development of our spectral multiscale method.

\section*{Acknowledgments}
The authors thank the referees for their suggestions.

%



\bibliographystyle{model1-num-names}

\begin{thebibliography}{00}


\bibitem {Kohlhoff}
S. Kohlhoff, P. Gumbsch, H.F. Fischmeister, Crack propagation in BCC crystals studied with a combined finite element and atomistic model, Philos. Mag. A $64$ ($1991$) $851$-$878$.

\bibitem {Abraham}
F.F. Abraham, J.Q. Broughton, N. Bernstein, E. Kaxiras, Spanning the continuum to quantum length scales in a dynamic simulation of brittle fracture, Europhys. Lett. $44$ ($1998$) $783$-$787$.

\bibitem {Broughton}
J.Q. Broughton, F.F. Abraham, N. Bernstein, E. Kaxiras, Concurrent coupling of length scales: Methodology and application, Phys. Rev. B $60$ ($1999$) $2391$-$2403$.

\bibitem {Rudd}
R.E. Rudd, J.Q. Broughton, Concurrent coupling of length scales in solid state systems, Phys. Stat. Sol. $217$ ($2000$) $251$-$291$.

\bibitem{RuddCGM}
R.E. Rudd and J. Q. Broughton, Coarse-grained molecular dynamics and the atomic limit of finite elements, Phys. Rev. B $58$ ($10$) ($1998$) $5893$-$5896$.

\bibitem{RuddCGM2}
R.E. Rudd and J. Q. Broughton, Coarse-grained molecular dynamics: nonlinear finite elements and finite temperature, Phys. Rev. B $72$ ($2005$) $144104$.

\bibitem{Belytschko}
S.P. Xiao, and T. Belytschko, A bridging domain method for coupling continua with molecular dynamics, Comput. Methods Appl. Mech. Eng. $193$ ($2004$) $1645$-$1669$.

\bibitem {Basu}
U. Basu, A. Chopra, Perfectly matched layers for time-harmonic elastodynamics of unbounded domains: theory and finite-element
implementation, Comput. Methods Appl. Mech. Engrg. $192$ ($2003$) $1337$-$1375$.

\bibitem {Tadmor}
E.B. Tadmor, M. Ortiz, R. Phillips, Quasicontinuum analysis of defects in solids, Philos. Mag. A $73$ ($1996$) $1529$-$1563$.

\bibitem {Tadmor2}
E.B. Tadmor, R. Phillips, M. Ortiz, Mixed atomistic and continuum models of deformation in solids, Langmuir $12$ ($19$) ($1996$) $4529$-$4534$.

\bibitem {Shenoy}
V.B. Shenoy, R. Miller, E.B. Tadmor, R. Phillips, M. Ortiz, Quasicontinuum models of interfacial structure and deformation, Phys. Rev. Lett. $80$ ($4$) ($1998$) $742$-$745$.

\bibitem {Miller}
R. Miller, M. Ortiz, R. Phillips, V. Shenoy, E.B. Tadmor, Quasicontinuum models of fracture and plasticity, Eng. Fract. Mech. $61$ ($3$–$4$) ($1998$) $427$-$444$.

\bibitem {MLuskin}
M. Luskin, C. Ortner, B. Van Koten, Formulation and optimization of energy-based blended quasicontinuum method, Comput. Methods Appl. Mech. Eng. $253$ ($2013$) $160$-$168$.

\bibitem {LMDTadmor}
L.M. Dupuy, E.B. Tadmor, R.E. Miller, R. Phillips, Finite-temperature quasicontinuum: molecular dynamics without all the atoms, Phys. Rev. Lett. $95$ ($2005$) $060202$.

\bibitem {ZTangAluru}
Z. Tang, H. Zhao, G. li, N.R. Aluru, Finite-temperature quasicontinuum method for multiscale analysis of silicon nanostructuires, Phys. Rev. B $74$ ($2006$) $064110$.

\bibitem {JMnGVenturini}
J. Marian, G. Venturini, B.L. Hansen, J. Knap, M. Ortiz, G.H. Campbell, Finite-temperature extension of the quasicontinuum method using Langevin dynamics: entropy losses and analysis of errors, Modelling Simul. Mater. Sci. $18$ ($2010$) $015003$.

\bibitem {WE2002}
W. E, B. Engquist, The heterogeneous multiscale methods, Commun. Math. Sci. $1$ ($2002$) $87$-$132$.

\bibitem {Xli2005}
X. Li, W. E, Multiscale modeling of the dynamics of solids at finite temperature, J. Mech. Phys. Solids $53$ ($2005$) $1650$-$1685$.

\bibitem {WE2007}
W. E, B. Engquist, X. Li, W. Ren, E. Vanden-Eijnden, Heterogeneous multiscale methods: a review, Commun. Comput. Phys. $2$ ($2007$) $367$-$450$.

\bibitem {Xli2010}
X. Li, J.Z. Yang, W. E, A multiscale coupling method for the modeling of dynamics of solids with application to brittle cracks, J. Comput. Phys. $229$ ($2010$) $3970$-$3987$.

\bibitem {WagnerBSM}
G.J. Wagner, W.K. Liu, Coupling of atomistic and continuum simulations using a bridging scale decomposition, J. Comput. Phys. $190$ ($2003$) $249$-$274$.

\bibitem {Liu}
W.K. Liu, E.G. Karpov, S. Zhang, H.S. Park, An introduction to computational nanomechanics and materials, Comput. Methods Appl. Mech. Engrg.
$193$ ($2004$) $1529$-$1578$.

\bibitem {Qian}
D. Qian, G.J. Wagner, W.K. Liu, A multiscale projection method for the analysis of carbon nanotubes, Comput. Methods Appl. Mech. Eng. $193$ ($2004$) $1603$-$1632$.

\bibitem {ParkMS}
 H.S. Park, E.G. Karpov, W.K. Liu, P.A. Klein, The bridging scale for two-dimensional atomistic/continuum coupling, Philos. Mag. A $85$ ($2005$) $79$-$113$.

\bibitem {TangBSM}
S. Tang, T.Y. Hou, W.K. Liu, A mathematical framework of the bridging scale method, International J. Numer. Methods Eng. $65$ ($2006$) $1688$-$1731$.

\bibitem {TangPSBSM}
S. Tang, T.Y. Hou, W.K. Liu, A pseudo-spectral multiscale metrhod: Interfacial conditions and coarse grid equations, J. Comput. Phys. $213$ ($2006$) $57$-$85$.

\bibitem {Farrell}
D.E. Farrell, H.S. Park, W.K. Liu, Implementation aspects of bridging scale method and application to intersonic crack propagation, International J. Numer. Methods Eng. $71$ ($2007$) $583$-$605$.

\bibitem {Chatfield}
C. Chatfield, The Analysis of Time Series, Chapman and Hall, ($1984$).

\bibitem {Doyle}
J.F. Doyle, A spectrally formulated finite element for longitudinal wave propagation, International Journal of Analytical and Experimental modal Analysis, $3$ ($1988$) $1$-$5$.

\bibitem {DoyleTN}
J.F. Doyle, T.N. Farris, A spectrally formulated finite element for flexural wave propagation in beams, International Journal of Analytical and Experimental modal Analysis, $5$ ($1990$) $13$-$23$.

\bibitem {DoyleBook}
J.F. Doyle, Wave Propagation in Structures, second ed., Springer Verlag, New York, $1997$.

\bibitem {Gopalakrishnan}
S. Gopalakrishnan, M. Martin, and J.F. Doyle, A matrix methodology for spectral analysis of wave propagation in multiple connected Timoshenko beams, Journal of Sound and Vibration, $158$ ($1992$) $11$-$24$.

\bibitem {Gopalakrishnan2}
S. Gopalakrishnan, J.F. Doyle, Wave propagation in connected waveguides of varying cross-section, Journal of Sound and Vibration, $175$ ($1994$) $347$-$363$.

\bibitem {Deepak}
B.P. Deepak, R. Ganguli, S. Gopalakrishnan, Dynamics of rotating composite beams: A comparative study between CNT reinforced polymer composite beams and laminated composite beams using spectral finite elements, Int. J. Mech. Sci., $64$ ($2012$) $110$-$126$.

\bibitem {Vinod}
K.G. Vinod, S. Gopalakrishnan, R. Ganguli, Free vibration and wave propagation analysis of uniform and tapered rotating beams using spectrally formulated finite elements, Int. J. Solids Struct, $44$ ($2007$) $5875$-$5893$.

\bibitem {GKrishnanBook}
S. Gopalakrishnan, A. Chakraborty, D. R. Mahapatra, Spectral Finite Element Method, Springer Verlag, London ($2008$).

\bibitem {MMitra}
M. Mitra, S. Gopalakrishnan, Spectrally formulated wavelet Fnite element for wave propagation and impact force identifcation in connected 1-D waveguides, Int. J. Solids Struct, $42$ ($2005$) $4695$-$4721$.

\bibitem {SGBook}
S. Gopalakrishnan, M. Mitra, Wavelet Methods for Dynamical Problems, CRC Press, Boca Raton, FL, $2010$.

\bibitem {Igawa}
H. Igawa, K. Komatsu, I. Yamaguchi, T. Kasai, Wave propagation analysis of frame structures using the spectral element method, Journal of Sound and Vibration, $277$ ($2003$) $1071$-$1081$.

\bibitem {Blais}
J.F. Blais, M. Cimmino, A. Ross, D. Granger, Suppression of time aliasing in the solution of the equations of motion of an impacted beam with partial constrained layer damping, Journal of Sound and Vibration, $326$ ($2009$) $870$-$882$.

\bibitem {Kishor}
D.K. Kishor, S. Gopalakrishnan, R. Ganguli, Wave propagation in acoustic fluids using spectral finite element model, Int. J. Numer. Meth. Fluids ($2010$) $1$-$53$.

\bibitem {Murthy}
M.V.V.S. Murthy, S. Gopalakrishnan, P.S. Nair, Signal wrap-around free spectral element formulation for multiply connected finite 1D waveguides, Journal of Aerospace science and Technologies, $63$ $1$ ($2011$) $72$-$88$.

\bibitem {Frenkel}
D. Frenkel, B. Smit, Understanding Molecular Simulation: From Algorithms to Application, second ed., Academic Press, New York, $2002$.

\bibitem {Adelman}
S.A. Adelman, J.D. Doll, Generalized Langevin equation approach for atom/solid-surface scattering: collinear atom/harmonic chain model, J. Chem. Phys. $61$ ($1974$) $4242$-$4245$.

\bibitem {Cai}
W. Cai, M. de Koning, V.V. Bulatov, S. Yip, Minimizing boundary reflections in coupled-domain simulations, Phys. Rev. Lett. $85$ ($13$) ($2000$) $3213$-$3216$.

\bibitem {EH}
W. E, Z.Y. Huang, Matching conditions in atomistic-continuum modeling of materials, Phys. Rev. Lett. $87$ ($13$) ($2001$) $135501$.

\bibitem {EH2}
W. E, Z. Huang, A dynamic atomistic-continuum method for simulation of crystalline materials, J. Comput. Phys. $182$ ($2002$) $234$-$261$.

\bibitem {WagnerMDBC}
G.J. Wagner, E.G. Karpov, W.K. Liu, Molecular dynamics boundary conditions for regular crystal lattice, Comput. Methods Appl. Mech. Engrg. $193$ ($2004$) $1579$-$1601$.

\bibitem {KarpovBC}
E.G. Karpov, G.J. Wagner, W.K. Liu, A Green's function approach to deriving non-reflecting boundary conditions in molecular dynamics, Int. J. Numer. Meth. Engng. $62$ ($2005$) $1250$-$1262$.

\bibitem {STangBC}
S. Tang, A finite difference velocity approach with velocity interfacial conditions for multiscale computations of crystatlline solids, J. Comput. Phys. $227$ ($2008$) $4038$-$4062$.

\bibitem {Xli2006}
X. Li, W. E, Variational boundary conditions for moleculardynamics simulations of solids at low temperature, Commun. Comput. Phys. $1$ ($2006$) $135$-$175$.

\bibitem {Xli2007}
X. Li, W. E, Variational boundary conditions for moleculardynamics simulations of crystalline solids at finite temperature: Treatment of the heat bath, Phys. Rev. B $76$ ($2007$) $104$-$107$.

\bibitem {Gonella}
S. Gonella, M. Ruzzene, Bridging scale analysis of wave propagation in heterogeneous structures with imperfections, Wave Motion, $45$ ($2008$) $481$-$497$.

\bibitem {Kittel}
C. Kittel, Introduction to Solid State Physics, seventh ed., John Wiley $\&$ Sons, Inc., U.K., $2008$.

\bibitem {GFriesecke}
G. Friesecke, K. Matthies, Geometric solitary waves in a $2$D mass-spring lattice, Discrete and Continuous Dynamical Systems $3$ ($1$) ($2003$) $105$-$114$.

\bibitem {SGopal}
S. Gopalakrishnan, A deep rod finite element for structural dynamics and wave propagation problems, Int. J. Numer. Meth. Eng.  $48$ ($2000$) $731$-$744$.


\bibitem {Weeks}
T.W. Weeks, Numerical inversion of Laplace transforms using Laguerre functions, Journal of the ACM, $13$ ($1966$) $419$-$429$.

\bibitem {Crump}
K.S. Crump, Numerical inversion of Laplace transforms using a Fourier series approximation, Journal of the ACM, $23$($1$) ($1978$) $89$-$96$.

\bibitem {Wilcox}
D.J. Wilcox, Numerical Laplace transformation and inversion, International Journal of Electrical Engineering Education, $15$ ($3$) ($1978$) $247$-$265$.

\bibitem {Wedepohl}
L.M. Wedepohl, Power systems transients: errors incurred in the numerical inversion of the Laplace transform, in: Proceedings of Midwest Symposiumon Circuits and Systems, Puebla, Mex,($1983$) $174$-$178$.

\bibitem {Ramirez}
A. Ramirez, P. Gomez, P. Moreno, A. Gutierrez, Frequency domain analysis of electromagnetic transients through the numerical Laplace transforms, in: Proceedings of IEEE Power Engineering Society of America, Denver,CO,USA, $1$ ($2004$) $1136$-$1139$.

\bibitem {Moreno}
P. Moreno, A. Ramirez, Implementation of numerical Laplace transform: A Review, IEEE Trans. Power Del., $23$ $4$ ($2008$) $2599$-$2609$.

\bibitem {Kreyszig}
E. Kreyszig, Advanced Engineering Mathematics, ninth ed., John Wiley $\&$ Sons, Singapore, $2006$.

\bibitem {ANag1}
A. Nag, D.R. Mahapatra, S. Gopalakrishnan, T.S. Sankar, A spectral finite element with embedded delamination for modeling of wave scattering in composite beams, Compos. Sci. Technology, $63$ ($2003$) $2187$-$2200$.

\bibitem {ANag2}
A. Nag, D.R. Mahapatra, S. Gopalakrishnan, Identification of delamination in a composite beam using a damaged spectral element, Struct. Health Monitoring, $1$ ($2002$) $105$-$126$.


%
%
%

\end{thebibliography}



%
\end{document}